\documentclass[aps,prd,10pt,notitlepage,showpacs,nofootinbib,superscriptaddress,twocolumn,raggedbottom,preprintnumbers,longbibliography]{revtex4-1}

\usepackage{amsfonts}
\usepackage{amsmath}
\usepackage{amssymb}
\usepackage{array}
\usepackage{bbold}
\usepackage{bm}
\usepackage{booktabs}
\usepackage{dsfont}
\usepackage{enumitem}
\usepackage[perpage]{footmisc}
\usepackage{array}
\usepackage{float}
\usepackage{graphicx}
\usepackage[utf8]{inputenc}
\usepackage{lastpage}
\usepackage{multirow}
\usepackage{nicefrac}
\usepackage[normalem]{ulem}
\usepackage[dvipsnames]{xcolor}
\usepackage{graphicx}
\usepackage{color}
\usepackage{bigstrut}
\usepackage{enumitem}
\usepackage{bm,mwe}
\usepackage[export]{adjustbox}
\usepackage{listings}
\usepackage{nameref}
\usepackage{dcolumn}
\usepackage{hyperref}
\hypersetup{
    colorlinks=true,
    urlcolor=blue,
    linkcolor = blue,
    urlcolor  = blue,
    citecolor = blue,
    anchorcolor = blue,
    pdftitle={a1 pheno},
    pdfauthor={friends},
    pdfsubject={a1}}

\graphicspath{{./plots/}}

\renewcommand{\Im}{{\rm Im}}

\newcommand*\dif{\mathop{}\!\mathrm{d}}
\newcommand\T{\rule{0pt}{2.6ex}}       
\newcommand\B{\rule[-1.2ex]{0pt}{0pt}} 
\newcolumntype{R}[2]{%
    >{\adjustbox{angle=#1,lap=\width-(#2)}\bgroup}%
    l%
    <{\egroup}%
}
\newcommand*\rot{\multicolumn{1}{R{90}{1em}}}

\begin{document}


\title{
Pole position of the \texorpdfstring{$a_1(1260)$}{a1(1260)} resonance in a three-body unitary framework
}

\author{Daniel Sadasivan}
\email{daniel.sadasivan@avemaria.edu}
\affiliation{Ave Maria University, Ave Maria, FL 34142, USA}
\author{Andrei~Alexandru}
\email{aalexan@gwu.edu}
\affiliation{The George Washington University, Washington, DC 20052, USA}
\affiliation{Department of Physics, University of Maryland, College Park, MD 20742, USA}
\author{Hakan~Akdag}
\email{akdag@hiskp.uni-bonn.de}
\affiliation{Helmholtz-Institut für Strahlen- und Kernphysik (Theorie) and
Bethe Center for Theoretical Physics, Universität Bonn, 53115 Bonn, Germany}
\author{Felipe Amorim}
\email{FelipeAugusto.deAmor@my.avemaria.edu}
\affiliation{Ave Maria University, Ave Maria, FL 34142, USA}
\author{Ruair\'{i}~Brett}
\email{rbrett@gwu.edu}
\affiliation{The George Washington University, Washington, DC 20052, USA}
\author{Chris~Culver}
\email{C.Culver@liverpool.ac.uk}
\affiliation{Department of Mathematical Sciences, University of Liverpool, Liverpool L69 7ZL, United Kingdom}
\author{Michael~D\"oring}
\email{doring@gwu.edu}
\affiliation{The George Washington University, Washington, DC 20052, USA}
\affiliation{Thomas Jefferson National Accelerator Facility, Newport News, VA 23606, USA}
\author{Frank~X.~Lee}
\email{fxlee@gwu.edu}
\affiliation{The George Washington University, Washington, DC 20052, USA}

\author{Maxim~Mai}
\email{maximmai@gwu.edu}
\affiliation{The George Washington University, Washington, DC 20052, USA}
\affiliation{Helmholtz-Institut für Strahlen- und Kernphysik (Theorie) and
Bethe Center for Theoretical Physics, Universität Bonn, 53115 Bonn, Germany}
%

\preprint{JLAB-THY-21-3533}

\begin{abstract}
Masses, widths, and branching ratios of hadronic resonances are quantified by their pole positions and residues with respect to transition amplitudes on the Riemann sheets of the complex energy-plane. In this study we discuss the analytic structure in the physical energy region of three-body scattering amplitudes on such manifolds. As an application, we determine the pole position of the $a_1(1260)$ meson from the ALEPH experiment by allowing for $\pi\rho$ coupled channels in S- and D-wave. We find it to be $\sqrt{s_0}=(1232^{+15+9}_{-0-11}-i266^{+0+15}_{-22-27})~\text{MeV}$.
\end{abstract}

\maketitle

\section{Introduction}
Hadronic resonances often decay strongly into three particles. Especially in the meson sector, three-body decays can be the dominant modes, e.g.~for axial mesons like the $a_1(1260)$~\cite{ParticleDataGroup:2020ssz}. Excited mesons are searched for in recent experimental efforts like GlueX~\cite{AlGhoul:2017nbp}, COMPASS~\cite{Alekseev:2009aa}, and at the  BESIII accelerator~\cite{Asner:2008nq} often in connection with exotic states that cannot consist of two constituent quarks only. For example, an exotic $\pi_1(1600)$ was found by COMPASS~\cite{Alekseev:2009aa} in three-pion decays. These experiments entail new partial-wave analysis (PWA) efforts, e.g.~by COMPASS~\cite{COMPASS:2015gxz, COMPASS:2018uzl}, BESIII~\cite{BESIII:2021aza, BES:2004twe}, CLEO~\cite{CLEO:1999rzk}, or in coupled channels using the PAWIAN framework for $p\bar p$ induced meson production~\cite{CrystalBarrel:2019zqh}. 

On the theory side, the final state interaction of three strongly interacting particles has been studied with Khuri-Treiman equations and similar frameworks by the Bonn group, JPAC, and others for light meson decays~\cite{Pasquier:1968zz, Aitchison:1976nk, Colangelo:2009db, Kubis:2009sb, Schneider:2010hs,Kampf:2011wr, Niecknig:2012sj, Guo:2011aa,Danilkin:2014cra, Guo:2014vya, Guo:2015zqa,  Daub:2015xja, Niecknig:2015ija, Guo:2016wsi, Isken:2017dkw, Albaladejo:2017hhj,  Niecknig:2017ylb, Dax:2018rvs, Jackura:2018xnx, Gasser:2018qtg,  Albaladejo:2019huw, JPAC:2019ufm, Mikhasenko:2019vhk, Akdag:2021efj}.
Faddeev-type arrangements of chiral two-body amplitudes were used to predict resonance states and study known ones~\cite{MartinezTorres:2007sr, Magalhaes:2011sh, MartinezTorres:2008kh, MartinezTorres:2011vh, Aoude:2018zty}. See also Ref.~\cite{Aitchison:2015jxa} for a pedagogical introduction into dispersive methods and Ref.~\cite{Aitchison:1966lpz} for connections between Khuri-Treiman equations and three-body unitary methods.

One such method applies the principle of three-body unitarity to construct  three-to-three amplitudes~\cite{Mai:2017vot}, extending earlier work~\cite{Aaron:1968aoz, Aaron:1973ca} to the above-threshold regime. The subthreshold behavior of this amplitude has been studied in Refs.~\cite{Jackura:2018xnx, Dawid:2020uhn} and new insights into covariant vs. time-ordered formulations for the interaction kernel were obtained recently~\cite{Zhang:2021hcl}. The amplitude of Ref.~\cite{Mai:2017vot} has been extended to formulate three-body resonance decays including a fit to the $a_1(1260)\to 3\pi$ lineshape and prediction of Dalitz plots~\cite{Sadasivan:2020syi}. This study is the basis of the current work.

Experimentally, the $a_1(1260)$ resonance can be produced in $\tau$-decays~\cite{CLEO:1999rzk,Schael:2005am} via ${\tau \rightarrow (3\pi) \nu_\tau}$. Therefore, its three-pion  dynamics can be separated off from the weak primordial interaction to be measured cleanly for the $I^G(J^{PC})=1^-(1^{++})$ quantum numbers. This  distinguishes this semileptonic $\tau$ decay from some of the aforementioned experiments in which multiple partial waves contribute to the final three-pion state. Of course, the $a_1$ resonance still couples to various configurations of the 2+1 pions, dominated by $\rho\pi$ in S-wave and $\sigma\pi$ in P-wave ($\sigma$ standing for the $f_0(500)$ resonance), but also several subdominant waves, see CLEO~\cite{CLEO:1999rzk}, COMPASS~\cite{COMPASS:2015gxz}, and  BESIII results~\cite{BESIII:2021aza}. Recent calculations based on chiral unitary methods predict that the $a_1(1260)\to\pi\sigma$ decay ratio is very small, in the few percent range~\cite{Molina:2021awn}. This is in contrast to an older phenomenological study~\cite{CLEO:1999rzk} finding a more substantial $\pi\sigma$ branching ratio.
This shows that, despite the clean experimental way to produce the $a_1(1260)$, its properties such as branching fractions are under continued debate. The resonance is very wide (with very large uncertainties  quoted by the PDG~\cite{ParticleDataGroup:2020ssz}), indicating strong and non-trivial three-body effects which makes it a prime candidate to study few-body dynamics. This is reflected in an increased interest in the properties and structure of the $a_1(1260)$~\cite{Janssen:1993nj, Lutz:2003fm, Roca:2005nm, Geng:2006yb, Wagner:2007wy, Wagner:2008gz, Lutz:2008km, Kamano:2011ih, Nagahiro:2011jn, Zhou:2014ila, Zhang:2018tko, JPAC:2018zwp, Sadasivan:2020syi, Dai:2020vfc, Dias:2021upl}, as well as the related $\tau$-decay~\cite{Bowler:1986exb, Kuhn:1992nz, Isgur:1988vm, Dumm:2009va, Nugent:2013hxa, Dai:2018zki}. 

The study of the $a_1(1260)$ with the \emph{ab-initio} techniques of lattice QCD has also made significant progress. For a pioneering calculation see Ref.~\cite{Lang:2014tia} where the $\rho$-meson was treated as a stable particle, motivated by the small box size. Recently, this approximation was lifted by using up to three pion operators in combination with the  finite-volume unitarity (FVU) three-body quantization condition~\cite{Mai:2017bge, Mai:2018djl} that allowed for the first pole extraction of a three-body resonance from lattice QCD~\cite{Mai:2021nul}. The infinite-volume version of that formalism is very similar to the one of Ref.~\cite{Sadasivan:2020syi} featuring coupled channels and explicit sub-channel ($\rho$) dynamics. See Refs.~\cite{Mai:2021lwb, Hansen:2019nir, Rusetsky:2019gyk} for reviews on recent progress of three-body physics in lattice QCD.

In this work, we use the formalism of Ref.~\cite{Mai:2021nul} to determine the $a_1(1260)$ pole position from experiment including statistical and some systematic uncertainties. This work is related to older determinations of the $a_1$ pole position~\cite{Janssen:1993nj} but also to Ref.~\cite{JPAC:2018zwp} (JPAC), in which the $S$-wave $\rho\pi$ channel was used to fit the $a_1$ lineshape~\cite{Schael:2005am} with an approximately unitary formalism. In contrast, our formalism is manifestly unitary, which considerably complicates the analytic structure through the pertinent pion exchange mechanism.  This requires a thorough discussion in Sec.~\ref{sec:analytic} based on the formalism summarized in Sec.~\ref{sec:formalism}. As such, it provides the only pole determination in three-body unitary amplitudes except for Ref.~\cite{Kamano:2011ih} and Ref.~\cite{Mai:2021nul}. However, in Ref.~\cite{Kamano:2011ih} the PDG pole position of the $a_1$ was fitted, while in this study we directly fit the lineshape from experiment. We therefore expect to extract the most reliable pole position of the $a_1(1260)$ resonance to date, with our results discussed in Sec.~\ref{sec:fit}.

\section{Formalism}
\label{sec:formalism}

The $a_1(1260)$ couples to  three-pion states in the $I^G(J^{PC})=1^-(1^{++})$ channel that can be decomposed as $\pi\rho$ in S/D-wave, $\pi \sigma$ and $\pi (\pi\pi)_{I=2}$ in P-waves and other channels. Phenomenologically $(\pi\rho)_{\rm S}$ is dominant~\cite{Kuhn:2004en} with the branching ratios into other channels quite uncertain~\cite{ParticleDataGroup:2020ssz}, see also Ref.~\cite{Molina:2021awn}. 
Therefore, we limit here the channel space to $\pi\rho$ in S and D  waves. Finally, we note that the isobar formulation of the two-body sub-channel dynamics used in this study is not an approximation but a re-parameterization of the full two-body amplitude~\cite{Bedaque:1999vb, Hammer:2017kms}.

 \begin{figure}
    \centering
    \includegraphics[width=\linewidth,trim=0cm 17cm 17cm 0cm,clip]{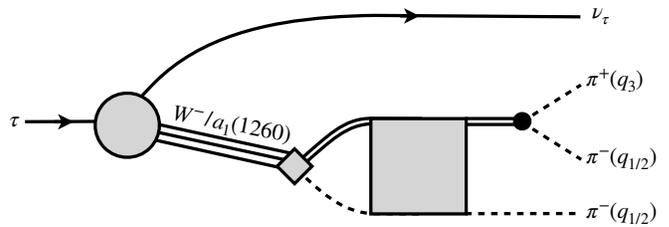}
    \caption{Factorization of the (weak) production mechanism and (hadronic) final state three-body interaction. Full-directed, dashed, and double-full lines denote leptons, mesons and auxiliary $\rho$ fields, respectively. The initial production mechanism is shown by a shaded circle and diamond corresponding to Eq.~\eqref{eq:firstdecayvertexprime}. The three-body unitary dynamics of the final pion states is depicted by the shaded rectangle, see Eq.~\eqref{eq:Gammabrev}.
    }
    \label{fig:process}
\end{figure}

Our formalism from Ref.~\cite{Sadasivan:2020syi} is summarized in the following. 
The $a_1$ lineshape with  $\pi^-\pi^-\pi^+$ final states, 
\begin{align}
\label{eq:lineshape}
&\mathcal{L}(\sqrt{s})=N(m_\tau^2-s)^2
\int \frac{\dif^3{\bm q}_1}{(2\pi)^3}
\frac{\dif^3{\bm q}_2}{(2\pi)^3}
\frac{\dif^3{\bm q}_3}{(2\pi)^3}
\frac{(2\pi)^4}{8E_{q_1}E_{q_2}E_{q_3}}
\nonumber \\
&
\times\delta^4(P_3-q_1-q_2-q_3)
\nonumber \\
&
\times
\left(
\big|\sum_\lambda\hat\Gamma_{-1\lambda}\big|^2
+\frac{m_\tau^2}{s}\big|\sum_\lambda\hat\Gamma_{0\lambda}\big|^2
+\big|\sum_\lambda\hat\Gamma_{+1\lambda}\big|^2
\right)
\end{align}
depends on the three-body energy $\sqrt{s}$ and scales with an irrelevant normalization $N$. Here, $q_1$, and $q_2$ are outgoing $\pi^-$ momenta that must be symmetrized later, $q_3$ is the outgoing $\pi^+$ momentum, $m_\tau$ is the mass of the $\tau$, and $E_x=\sqrt{\bm{x}^2+m_\pi^2}$ here and in the following. 
The term $(m_\tau^2-s)^2$ accounts for the $\tau\rightarrow W^-\nu_\tau$ decay vertex and the two-body phase space of the $a_1$ and the $\nu_\tau$ of this process after integration over the neutrino angles~\cite{JPAC:2018zwp}. See Fig.~\ref{fig:process} for a graphical representation of the complete $\tau$ decay process. Furthermore, we chose the total four-momentum of the three-body system $P_3=(\sqrt{s},{\bf 0})$. 

The amplitude $\hat\Gamma_{\varLambda\lambda}\equiv\hat\Gamma_{\varLambda\lambda}(\boldsymbol{q}_1,\boldsymbol{q}_2,\boldsymbol{q}_3)$ describes the decay of the axial $a_1(1260)$ resonance at rest with helicity $\varLambda$ measured along the $z$-axis into a $\pi^-$ and a $\rho_\lambda^0\to \pi^+\pi^-$ with helicity $\lambda$,
\begin{align}
\hat \Gamma_{\varLambda\lambda} (\boldsymbol{q}_1,\boldsymbol{q}_2,\boldsymbol{q}_3)
&=\frac{1}{\sqrt{2}}\big[\Gamma_{\varLambda\lambda} (\boldsymbol{q}_1,\boldsymbol{q}_2,\boldsymbol{q}_3)
-\left(\boldsymbol{q}_1\leftrightarrow\boldsymbol{q}_2\right)\big],
\end{align}
where the minus sign in the exchange term comes from the overall odd intrinsic parity of the process,
\begin{align}
\label{eq:Gamma}
\Gamma_{\varLambda\lambda} ({\bm q}_1,{\bm q}_2,{\bm q}_3)&=
\sqrt{\frac{3}{4\pi}}\, {\mathfrak D}^{1*}_{\Lambda\lambda}(\phi_1,\theta_1,0)\times
\\\nonumber 
&\hspace{6em} 
v^\pm_{\lambda}(q_2,q_3)U_{\lambda L}\breve\Gamma_{L}(q_1) \,,
\end{align}
and
\begin{align}
\label{eq:Gammabrev}
\breve{\Gamma}_{L} (q_1)
&=\tau(\sigma(q_1))\bigg[\,D_{L}(q_1)+ \\
\nonumber
&\hspace{4em}
\int\limits_0^\Lambda \frac{\dif p \, p^2}{(2\pi)^3}\,\frac{1}{2E_p}\,
T^{J}_{LL'}(q_1,p)
\tau(\sigma(p))
D_{L'}(p)\bigg]\,.
\end{align}
For readability (confusion with four-vector notation is excluded by context), we have abbreviated
$D(x):=D(|\bm{x}|)$, $\breve{\Gamma}(x)=\breve{\Gamma}(|\bm{x}|)$,
$T(q_1,p):=T(|\bm{q}_1|,|\bm{p}|)$, and $\sigma(x):=\sigma(|\bm{x}|)$ where the two-body invariant mass squared is denoted by
\begin{align}
\sigma(x)=s+m_\pi^2-2\sqrt{s}E_x \,.
\label{eq:sigma}
\end{align}
The angular structure of the final $\pi\rho$ state is conveyed by the usual capital Wigner-D function, $\mathfrak{D}^{J}_{\Lambda\lambda}(\phi_1,\theta_1,0)$
 with angles $\theta_1$ and $\phi_1$ giving the polar and azimuthal angles of $\bm{q}_1$. 
Note that the third argument is set to zero in the current convention, cf.~Ref.~\cite{Berman:1965gi}, which is consistent with the $\rho$ polarization vectors of Appendix~\ref{app:a} obtained through a boost and two rotations (no initial rotation about the $z$-axis).

Equation~\eqref{eq:Gamma}  contains the transformation from the $JLS$ basis to the helicity basis, with $L$ denoting the orbital angular momentum between $\pi$ and $\rho$ and $J=S=1$ for total and $\rho$ spin, respectively.  This transformation involves the matrix 
\begin{align}
U_{L\lambda }&=\sqrt{\frac{2L+1}{2J+1}}(L01\lambda|J\lambda)
(1\lambda00|1\lambda) \ ,\nonumber \\
U&
=\left(
\begin{array}{ccc}
 \frac{1}{\sqrt{3}} & \frac{1}{\sqrt{3}} & \frac{1}{\sqrt{3}}\\
 \frac{1}{\sqrt{6}} & -\sqrt{\frac{2}{3}} & \frac{1}{\sqrt{6}}\\
\end{array}
\right)
\,,
\label{eq:umat}
\end{align}
expressed by Clebsch-Gordan coefficients~\cite{Chung:1971ri}, and ${U_{\lambda L}=U_{L\lambda}}$, while we sum over identical indices $L$ and  $L'$ in Eqs.~(\ref{eq:Gamma}) and \eqref{eq:Gammabrev}, respectively.

The final decay vertex $v^\pm$ for $\rho^0\to\pi^+\pi^-$ in Eq.~\eqref{eq:Gamma} reads 
\begin{align}
v^{\pm}_{\lambda}(q_2,q_3)&=I'v_{\lambda}(q_2,q_3)\,,\\
v_{\lambda}(q_2,q_3)&=
-ig_1\epsilon_{\lambda}^\mu(\boldsymbol{q}_1)~(q_2-q_3)_\mu 
\,,
\label{eq:vs}
\end{align}
where $q_2$, $q_3$ denote four-momenta, $\boldsymbol{q}_1=-\boldsymbol{q}_2-\boldsymbol{q}_3$, $g_1$  is the $\rho\to\pi\pi$ coupling, $v_\lambda$ is the isospin-1 projected decay vertex, and $I'$ describes the transition from isospin to particle basis as needed only in the final $\rho$ decay. Note that the latter factor is irrelevant as long as there is only one isobar ($\rho^0$). Then, this factor can be reabsorbed into the overall normalization of the $a_1$ decay.

Continuing with the description of Eq.~\eqref{eq:Gammabrev},
the $\rho$ propagator $\tau$ is discussed in more detail in Sec.~\ref{sec:twobody}.
Furthermore, the $a_1\to \rho \pi$ vertex, $D$, in Eq.~\eqref{eq:Gammabrev} is directly parameterized in the $JLS$ basis as
\begin{align}
\label{eq:firstdecayvertexprime}
D_{L'}(p)=\left(D_{fL'}
+\frac{m^2_\pi \sqrt{c^{(-1)}_{L'L'}} D_{\tilde{f}} }{s-m^2_{a_1}}\right)\left(\frac{p}{m_\pi}\right)^{L'} \,,
\end{align}
where $D_{fL'}$ for $L'=0,2$ and $D_{\tilde{f}}$ are free parameters that are fit to the lineshape accounting, as well, for its unknown overall normalization.
The quantity $m_{a_1}$ is the expansion point of the three-body force and $c^{(-1)}_{L'L'}>0$ is an expansion parameter (see below). Its square root may be understood as a bare $a_1\pi\rho$ coupling.

The quantity $T^J_{LL'}$ in Eq.~\eqref{eq:Gammabrev} is the isobar-spectator amplitude in the $JLS$ basis given by 
\begin{align}
\label{eq:Bethe-SalpeterPartial}
&T^J_{LL'}(q_1,p)=\left(B^J_{LL'}(q_1,p)+C_{LL'}(q_1,p)\right)+
\\
&\int\limits_0^\Lambda \frac{\dif l\, l^2}{(2\pi)^3 2E_l} \left(B^J_{LL''}(q_1,l)+C_{LL''}(q_1,l)\right)\nonumber
\tau(\sigma(l)) T^J_{L''L'}(l,p)\nonumber
\,,
\end{align}

where summation over $L''$ is implied.
Note that the indices correspond to matrix notation, i.e.~the first indices $L$ and $q_1$ label outgoing (angular) momentum while the second indices $p$ and $L'$ label incoming  (angular) momentum (similarly, in Eqs.~\eqref{eq:Gamma} and \eqref{eq:Gammabrev}).
The integrations in Eq.~\eqref{eq:Gammabrev} and \eqref{eq:Bethe-SalpeterPartial} have been regularized by the same cutoff $\Lambda$ in contrast to Ref.~\cite{Sadasivan:2020syi} where covariant form factors were used. We prefer here a hard cutoff because it simplifies the analytic continuation as discussed in Sec.~\ref{sec:analytic}  which also contains the in-depth description of the contours for the integrations in Eqs.~\eqref{eq:Gammabrev} and \eqref{eq:Bethe-SalpeterPartial}. 

In Eq.~\eqref{eq:Bethe-SalpeterPartial} the $\pi\rho$ interaction term $B$ is complex-valued as demanded by three-body unitarity~\cite{Mai:2017vot} and obtained from the plane-wave expression in isospin $I=1$,
\begin{align}
B_{\lambda\lambda'}(\bm{p},\bm{p}')=\frac{ v^*_{\lambda}(P-p-p',p) v_{\lambda'}(P-p-p',p')}{2E_{p'+p}(\sqrt{s}-E_p-E_{p'}-E_{p'+p}+i\epsilon)}\,,
\label{eq:Bterm}
\end{align}
by projecting it to angular momenta ${L^{(\prime)} \in\{S,D\}}$ via
\begin{align}
\label{eq:BmatrixPartial}
B^J_{\lambda\lambda'}(q_1,p)=
2\pi\int\limits_{-1}^{+1} \dif x \, d^J_{\lambda\lambda'}(x) B_{\lambda\lambda'}(&{\bm q}_1,{\bm p}) \,,
\end{align}
 where  $d^J_{\lambda\lambda'}(x=\cos{\theta})$ denotes the small Wigner-d function and $\theta$ is the $\pi\rho$ scattering angle.
Subsequently, the $JLS$ expression is obtained by a linear transformation,
\begin{align}
\label{eq:transformation}
B^J_{LL'}(q_1,p)=U_{L\lambda}B^J_{\lambda\lambda'}(q_1,p) U_{\lambda' L'}\,,
\end{align}
with $U_{L\lambda}$ from Eq.~\eqref{eq:umat} and, as before, $\lambda',\,L',\,p$ ($\lambda,\,L,\,q_1$) label the incoming (outgoing) state.

Three-body unitarity allows for additional terms of the $\pi\rho$ interaction that need to be real in the physical region~\cite{Mai:2017vot}. We refer to such terms as contact terms or
three-body forces that are generically parameterized by a Laurent series in the $JLS$ basis (${L^{(\prime)} \in\{\text{S,D}\}}$),
\begin{align}
C_{LL'}(p,p')&=\sum_{i=-1}^\infty c_{LL'}^{(i)}\left(\frac{s-m_{a_1}^2}{m_\pi^2}\right)^i
\frac{p^L\,p^{\prime L'}}{m_\pi^{L+L'}}
\,,
\label{eq:C-term}
\end{align}
including first-order poles to account for explicit resonances. 

We fit the parameters $m_{a_1}$, $c_{00}^{(-1)}$, and $c_{00}^{(0)}$ with all other parameters set to zero, meaning that the $C$-term couples directly to the S$\,\to\,$S-wave transition but only indirectly to D-wave through the $B$-term~\eqref{eq:transformation}. The analysis shows that this restriction is sufficient to fit the lineshape data as discussed in Sec.~\ref{sec:fit}. Of course, in future fits to Dalitz plots the data become more sensitive to the partial-wave content and we expect that more fit parameters and channels are needed. 

To fit the $a_1$ lineshape one needs to continue $\breve{\Gamma}$ to real spectator momenta and perform the phase space integration over the final three-pion state. This is described in detail in Ref.~\cite{Sadasivan:2020syi} but is not repeated here.

\begin{figure}
\centering

    \includegraphics[width=\linewidth]{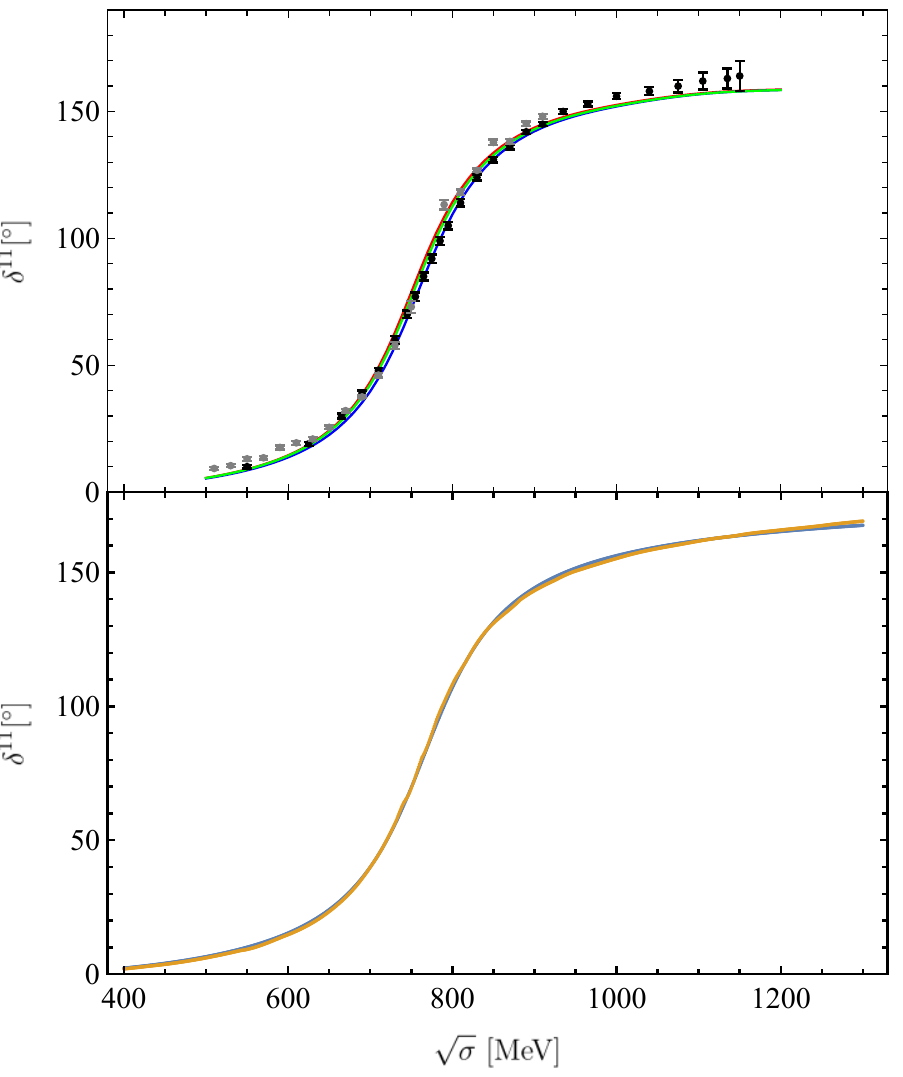}
    \caption{Upper Plot: The phaseshift from Eq.~\eqref{eq:phaseshift} with  free parameters $a_0$ and $a_1$ fitted to data from Refs.~\cite{Protopopescu:1973sh} (black circles) and \cite{Estabrooks:1974vu} (gray circles) is indicated in green (case (a)). The fit to only the data of Ref.~\cite{Protopopescu:1973sh}, case (b), is indicated in blue (to \cite{Estabrooks:1974vu} only, case (c): in red). Lower Plot, case (d): The dispersive solution from Ref.~\cite{Pelaez:2019eqa} (orange), and the three-parameter fit to it (purple).
   }
    \label{fig:phaseshift}
\end{figure}

\subsection{Two-body input}
\label{sec:twobody}

The $\rho$ propagator $\tau$ of Eq.~\eqref{eq:Gammabrev} in $n$-times subtratced form reads
\begin{align}
\tau^{-1}(\sigma)&=K_n^{-1}(\sigma)
-\Sigma(\sigma) \,,\nonumber \\
\Sigma(\sigma) &=\int\limits_0^\infty 
\frac{\dif k\,k^2}{(2\pi)^3}\,\frac{1}{2E_k}\,\left(\frac{\sigma}{\sigma^{\prime}}\right)^n
\,\frac{\tilde v^2(k)}{\sigma-4E_k^2+i\epsilon} \,,
\nonumber \\
\sigma^\prime&=(2E_k)^2 \,,~~ 
\tilde v(k)=\sqrt{\frac{16\pi}{3}}\,g_1 k
\label{eq:tau-infinite}
\end{align}
with a self-energy $\Sigma$, that converges for $n\ge 2$, and a regular $K$-matrix like quantity.
The former is expressed in terms of the vertex $\tilde v$ projected to the ${I=1}$ P-wave (spin $S=1$) quantum numbers. This vertex can be obtained by considering the first-order Born series for the $\pi(k_1)\pi(k_2)\to\pi(k_1')\pi(k_2')$ scattering amplitude in the two-body rest frame, i.e.~$k_1^{(\prime)}+k_2^{(\prime)}=(\sqrt{\sigma},\bm{0})$, ${k_1^{(\prime)\mu}-k_2^{(\prime)\mu}=(0,2\bm{k}^{(\prime)})}$. Using the $\rho\pi\pi$ vertex from Eq.~\eqref{eq:vs}
this reads
\begin{align}
T_\rho(\sigma,z)&=I_sI_\rho\,\frac{\sum_\lambda v_\lambda(k_1,k_2)v^*_{\lambda}(k_1^{\prime},k_2^{\prime})}{\sigma-m_\rho^2}
=\frac{g_1^2}{\sigma-m_\rho^2}\,4kk'z \,,
\label{eq:rhos}
\end{align}
where $z=\bm{k}\cdot\bm{k}'/(kk')$, $I_\rho=2$ is a factor for isospin-1, and $I_s=\nicefrac{1}{2}$ is a symmetry factor.  The second equal sign in Eq.~\eqref{eq:rhos} is due to the general properties of the helicity state vectors, cf.~Eq.~\eqref{eq:helicity-sum}. Projecting this amplitude to the P-wave amounts then to
\begin{align}
T_\rho^1(\sigma)
&=2\pi\int\limits_{-1}^1\dif z\,P_1(z)\, T_\rho(\sigma,z) \\
&=\left(\sqrt{\frac{16\pi}{3}}\,g_1k\right)\frac{1}{\sigma-m_\rho^2}
\left(\sqrt{\frac{16\pi}{3}}\,g_1k'\right)\nonumber\\
&=:\frac{\tilde v(k)\tilde v(k')}{\sigma-m_\rho^2} \nonumber\,,
\end{align}
defining the projected vertex $\tilde v$ that automatically fulfills $\tilde v=\tilde v^*$.

The two-body dynamics is encoded in $\tilde v$ but also in $K$ of Eq.~\eqref{eq:tau-infinite} that is very similar to a $K$-matrix, up to the selfenergy $\Sigma$ that contains also a real part. The subtraction polynomial $K$ reads
\begin{align}
K_n^{-1}(\sigma)=m_\pi^2\sum_{i=0}^{n-1} a_i \left(\frac{\sigma}{m_\pi^2}\right)^i  \,.
\label{eq:Kmatrix}
\end{align}

The parameters $a_i$ are fitted to the $\pi\pi$ phaseshift data by introducing the two-to-two on-shell $T$-matrix for $I=S=1$, 
 \begin{align}
T_{22}(\sigma)=\tilde v(k_\text{cm})\tau(\sigma)\tilde v(k_\text{cm}) \ ,
\label{eq:t22}
\end{align}
where  $k_\text{cm}=\sqrt{\sigma/4-m_\pi^2}$. The connection to the vector, isovector phaseshift $\delta^{11}$ is given by
\begin{align}
\delta^{11}(\sigma)=\tan^{-1}\left(\frac{\text{Im}\,T_{22}(\sigma)}{\text{Re}\,T_{22}(\sigma)}\right) \,,
\label{eq:phaseshift}
\end{align}
which depends on $g_1$ from Eq.~\eqref{eq:vs} and the $a_i$ from Eq.~\eqref{eq:Kmatrix}. However, $g_1$ is fully correlated with the $a_i$ so we fix it at $g_1=1$. 

To assess systematic effects from the two-body input we perform different fits: {\bf (a)} In a twice-subtracted fit ($n=2$), data from both experimental phaseshifts from Refs.~\cite{Protopopescu:1973sh, Estabrooks:1974vu} 
are fitted as shown in Fig.~\ref{fig:phaseshift} with the green line. As can be seen in the figure, the two sets of data are not in perfect agreement. In order to account for this source of systematic uncertainty we perform two additional fits: {\bf (b)} only to data from Ref.~\cite{Protopopescu:1973sh}, and {\bf (c)} only to the data of Ref.~\cite{Estabrooks:1974vu}. 
{\bf (d)}
Additionally, it has been shown in Refs.~\cite{Ananthanarayan:2000ht,Colangelo:2001df,Garcia-Martin:2011iqs,Colangelo:2018mtw,Pelaez:2019eqa} that these data have certain inconsistencies and that by imposing S-matrix principles such as crossing  symmetry the $\rho$ phase shift can be improved. Therefore, we perform an additional fit to the recent phase-shift parameterization from Ref.~\cite{Pelaez:2019eqa} using three subtractions $(n=3)$.

The parameters and the corresponding $\rho$ pole positions for cases (a) to (d) are given in  Table~\ref{tab:two-body-parameters}. 
The pole position for case (d) is  close to the one of Ref.~\cite{Pelaez:2019eqa} itself, of about $\sqrt{\sigma_\rho}=(763\pm 1.5-i(73\pm 1.5))$~MeV. 
As the table shows the $\rho$ pole positions for cases (a) to (d) are further apart than the statistical uncertainties. The latter was calculated for case (a) from resampling the phase shift data as indicated in the table. 
We will propagate these statistical errors and also the systematic differences between cases (a) to (d) to the three-body sector as described in Sec.~\ref{sec:discussion}.

\begin{table}[t]
\begin{ruledtabular}
\begin{tabular}{lllll}
Input: & (a)\cite{Protopopescu:1973sh,Estabrooks:1974vu} & (b)\cite{Protopopescu:1973sh}  & (c)\cite{Estabrooks:1974vu} & (d)\cite{Pelaez:2019eqa} \B\\
\hline
$a_0$&$-0.460(2)$  &$-0.471$ & $ -0.464$ & $-0.327$  \T
\\
$a_1\cdot 10$&$+0.156(1)$ & $+0.157$ & $+0.157$ & $0.062$\B \\
$a_2\cdot 10^3$ & - & - & - & 0.259\\
Re~$\sqrt{\sigma_\rho}$ [MeV]&$754 (< 1)$& $758$ & $753$ & 766 \\
Im~$\sqrt{\sigma_\rho}$ [MeV]&$-72(<1)$ & $-71$ & $-71$ & -74 
\end{tabular}\end{ruledtabular}
\caption{Fitted parameters of the two-body subsystem according to Eq.~\eqref{eq:Kmatrix} and $\rho(770)$ meson pole positions $\sqrt{\sigma_\rho}$. Statistical uncertainties are only quoted for the combined fit to both data sets and show that  systematic effects are larger (i.e., fits to the individual data sets).
\label{tab:two-body-parameters}}
\end{table}
\begin{figure}[tbh]
    \centering
    \includegraphics[width=0.9\linewidth]{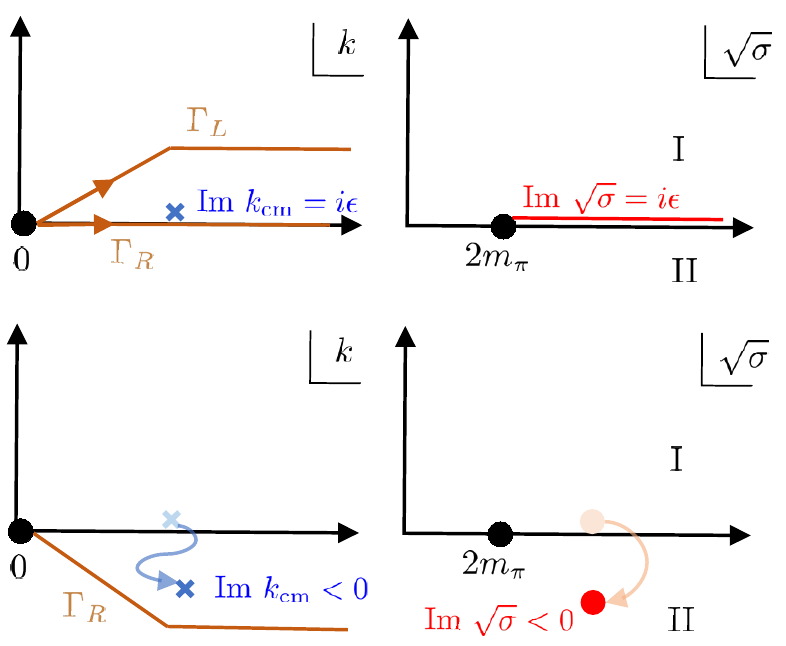}
    \caption{Contours in two-body scattering. Upper row: Momentum integration (left)  and physical amplitude (right) at $\sqrt{\sigma}+i\epsilon$ on the first (``I'') Riemann sheet. Lower row: Momentum integration and amplitude on unphysical sheet II with Im~$\sqrt{\sigma}<0$. See text for additional explanations. 
    }
    \label{fig:two-body_schematic}
\end{figure}

\section{Analytic continuation}
\label{sec:analytic}

The key ingredient of the discussed $a_1$-production mechanism is the integral equation~\eqref{eq:Bethe-SalpeterPartial} solved for the $\pi\rho$ scattering amplitude. This equation is solved by replacing the integrations over the real-valued magnitude of the meson momenta with complex values along certain contours described in the following. 
Additionally, and in view of the final goal of this study -- determination of the $a_1(1260)$ resonance pole -- one needs to analytically continue the scattering amplitude~\eqref{eq:Bethe-SalpeterPartial} in the three-body energy $\sqrt{s}$ to complex values. 

Specifically, two types of integrations occur: (1) in $l:=|\bm{l}|$ within the integral equation~\eqref{eq:Bethe-SalpeterPartial}; and (2) in $k:=|\bm{k}|$ within the selfenergy term of the two-body subsystem~\eqref{eq:tau-infinite}. The corresponding complex contours can be chosen individually and are referred to in the following as `spectator momentum contour' (SMC) and `selfenergy contour' (SEC), respectively.
Both contours start at the respective origins, $l=k=0$, and end at $l=\Lambda$ and $k=\infty$, respectively. In between these limits, different choices for the contours  define different Riemann sheets in $\sqrt{s}$ as discussed in the following. 

A similar discussion of the analytic structure in the context of dynamical coupled-channel approaches can be found in Ref.~\cite{Doring:2009yv} for the Jülich/Bonn/Washington approach~\cite{Ronchen:2012eg, Ronchen:2014cna, Mai:2021vsw} and in Ref.~\cite{Suzuki:2010yn} for the EBAC/ANL-Osaka approach~\cite{Julia-Diaz:2009dnz, Kamano:2013iva}. There is also a discussion in Ref.~\cite{JPAC:2018zwp} on analytic continuation, but the structure of the scattering equation is substantially different because it does not contain three-body cuts from pion exchange as demanded by three-body unitarity~\cite{Mai:2017vot}. In Ref.~\cite{Doring:2009yv}, a continuation obtained by certain approximations for three-body cuts was discussed, but the method proposed here is rigorous. Regularization is often achieved with form factors, and the SMC is given by a straight line from $l=0$ into the lower complex-momentum half-plane; see, e.g.~Refs.~\cite{Ronchen:2012eg, Huang:2011as, Ronchen:2018ury,Sadasivan:2020syi}. We refrain from the use of form factors because they make the analytic structure of the amplitude unnecessarily complicated and use a cutoff $\Lambda$ instead.

\begin{figure}[tbh]
    \centering
    \includegraphics[width=1.\linewidth]{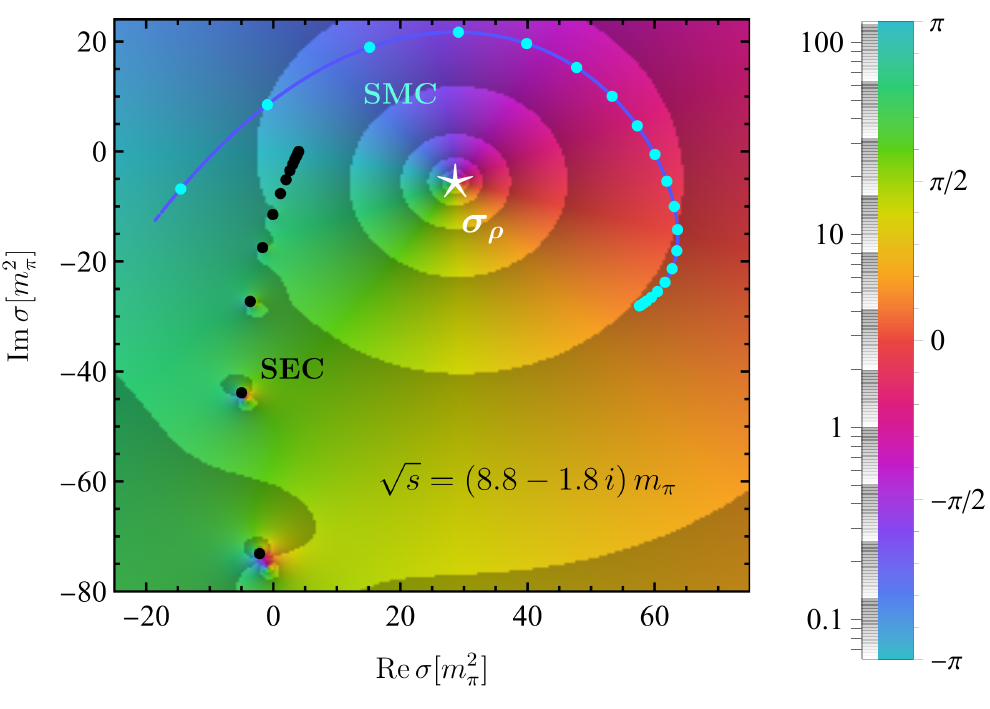}
    \caption{Example for SMC and SEC contours in the complex plane of the two-body energy-squared, $\sigma$. Typical Gauss node distributions on the contours are indicated with turquoise and black dots, respectively. The shading and coloring indicate magnitude and phase $\phi$ of $\tau=|\tau|e^{i\phi}$, respectively. The $\rho$ pole at $\sigma=\sigma_\rho$ is highlighted with the white star. See text for further explanations.
    }
    \label{fig:integrations}
\end{figure}

\subsection{Two-body scattering}
\label{sec:two-body}

To discuss the analytic structure, we recall that the 
placement of cuts is a choice, and that only the branch point marking the energy at which a cut begins is fixed. Cuts are the curves along which different Riemann sheets are analytically ``glued'' together. For example, a common choice in two-body scattering is to run the physical, right-hand cut along the real axis from threshold to $\infty$ in the two-body energy-squared $\sigma$. This simplifies the formulation of the dispersion relations and provides a convenient definition of the first and second Riemann sheet. With that definition, resonance poles can only  be found on the second, ``unphysical'' Riemann sheets as demanded by causality (see, e.g.~Ref.~\cite{gribov2008strong} for a proof). 

In contrast to cuts, the position of a branch point is fixed in $s$ and $\sigma$ for three- and two-particle scattering, respectively. Branch points define thresholds and arise whenever the pertinent momentum integrations begin in singularities or branch points themselves~\cite{Doring:2009yv}. This is illustrated in Fig.~\ref{fig:two-body_schematic} for the two-body case above threshold. The placement of the SEC producing the physical amplitude is constrained by the $+i\epsilon$ term in Eq.~\eqref{eq:tau-infinite}. 
In the figure, two possible integration contours are depicted, passing the singularity at $k=k_\text{cm}+i\epsilon$ either on the right ($\Gamma_R$) or left ($\Gamma_L$). The former does not change the sign of ($\Im\,\Sigma$) and, thus, yields the physical amplitude~\eqref{eq:t22} describing experimental measurements at real energies. In contrast, choosing $\Gamma_L$ leads to a sign change in $(\Im\,\Sigma)$ and an unphysical $T_{22}$ on sheet II (still, at the same ${\sigma=\sigma_\text{phys}+i\epsilon}$).

The physical and unphysical scattering amplitudes $T_{22}$ are connected to each other, smoothly building the $2^N$ Riemann sheets ($N$ being the number of two-body thresholds). By convention, the physical amplitude on  sheet I in the upper half-plane of $\sigma\in\mathds{C}$ is connected along the real axis, $\sigma\in [4m_\pi^2,\infty)$, to the unphysical sheet II in the lower half-plane.
 For energies with Im~$\sigma<0$ (see Fig.~\ref{fig:two-body_schematic}, lower right), the  amplitude  on sheet II can be obtained by deforming the SEC as shown to the lower left. In particular, the two-body singularity at $k_\text{cm}=\sqrt{\sigma/4-m_\pi^2}$ also acquires a negative imaginary part, but a smooth deformation of $\Gamma_R$ ensures that the SEC still passes the two-body singularity to the right.  This guarantees that the amplitude has been analytically continued from physical scattering energies $\sigma$ to the second sheet in the lower half-plane, where resonance poles can be found.

In summary, passing the two-body singularity to the left or to the right ($\Gamma_L$ vs. $\Gamma_R$) defines the Riemann sheet, except for one point at $\sigma=4m_\pi^2$. There, the two-body singularity coincides with the lower limit of the integration, $k=k_\text{cm}=0$. Consequently, at this point there is no distinction between sheets, i.e.~one is at the branch point that defines the two-pion threshold. We stress the (otherwise trivial) fact that a singularity in an integration limit induces a branch point, because it helps identifying branch points for the more complicated three-body case discussed in Sec.~\ref{sec:complexbp}; see also Ref.~\cite{Doring:2009yv}.

In regards to the present application to the $a_1(1260)$ channel, we chose the SEC for the $\pi\pi$ subsystem in the $\rho$ channel as
\begin{align}
\Gamma_\text{SEC}=\left\{k|k=t+\frac{ic_1}{2}\,\arctan(c_2t),t\in[0,\infty)
\right\} \,,
\label{eq:SEC}
\end{align}
with shape parameters $c_1$ and $c_2$ chosen such that this contour lies in the lower right quadrant
of the $k$ plane and always avoids the two-body singularity except at threshold, 
\begin{align}
\Gamma_\text{SEC}\cap\{k_\text{cm}\}\backslash\{0\}=\O \,.
\label{eq:cond0}
\end{align}
To display the SEC and the $\rho(770)$ resonance pole in the same plot, $\Gamma_\text{SEC}$ is mapped to the $\sigma$ plane according to $\sigma=4(m_\pi^2+k^2)$. The result is labeled ``SEC'' with the black circles in Fig.~\ref{fig:integrations} indicating the  Gauss nodes chosen for numerical integration. As the figure shows, the chosen $\Gamma^\prime_R$ is sufficiently deformed to not only allow for the calculation of the physical amplitude but also for the calculation of $\tau$ in a large portion of unphysical sheet II, bound by the mapped SEC. This portion includes the $\rho$ pole (white star). In other words, the so-defined two-body amplitude has its actual cut along the mapped SEC. Furthermore, instead of a discontinuity in $\sigma$ along that cut, the amplitude exhibits a series of poles which is a consequence of the discretization in a finite number of Gauss nodes. This is made visible in the figure through the shading (repeated transitions from transparent to dark gray indicating increasing values of $|\tau|$), in addition to the color coding that indicates the phase $\phi$ of $\tau=|\tau|e^{i\phi}$. Notably, the $\rho$ pole exhibits one full cycle $-\pi\to\phi\to+\pi$ (blue $\to$ red $\to$ green), indicating that the $\rho$ pole is indeed a first-order singularity as required for a resonance.

The idea of suitably constructing contours to access the Riemann sheet(s) of interest, where resonance poles are situated, can be generalized to three-body scattering as discussed in the following. In particular, contour deformation replaces other methods of analytic continuation in which explicit discontinuities have to be added to the amplitude.

\begin{figure}[t]
    \centering
    \includegraphics[width=1.\linewidth]{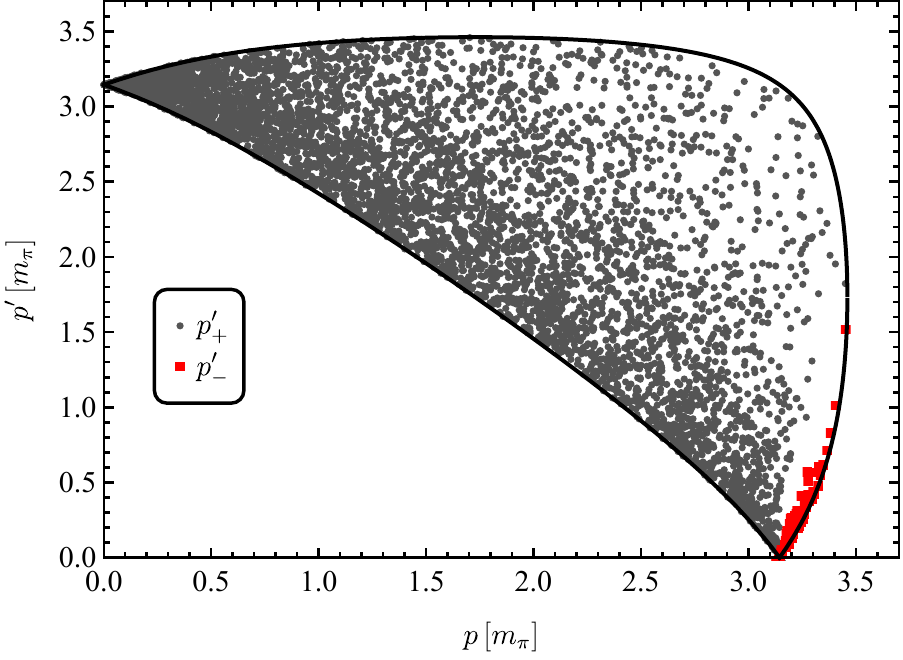}
    \caption{Domain of real solutions $p_\pm'(p,x)$ according to  Eq.~\eqref{eq:Bsingularities} for $\sqrt{s}=7.6\,m_\pi$. Gray and red dots represent individual solutions for some $x\in [-1,+1]$ which are enclosed by the $p_\pm'(p,x=\pm 1)$ boundary.
    }
    \label{fig:physmom}
\end{figure}

\subsection{Three-body cuts}
\label{sec:3bcuts}

Before turning to the construction of a suitable spectator momentum contour (SMC) to access the $a_1(1260)$ pole, one needs to discuss three-body cuts because they must be avoided by the SMC. This has been known for a long time and is discussed in the context of the $a_1$ resonance in Ref.~\cite{Janssen:1994uf} (we adapt and extend the discussion here). These cuts arise from the pion-exchange term of Eq.~\eqref{eq:Bterm} that is a direct consequence of three-body unitarity~\cite{Mai:2017vot}. We  note that this term corresponds to the forward-going part of pion exchange only. If one adds the backward-going part, one recovers the covariant denominator $u-m_\pi^2+i\epsilon$~\cite{Mai:2017vot} but we refrain from using this term as it can induce unphysical unitarity-violating imaginary parts above threshold if the regularization is not chosen correctly. Note also the related but different discussions on sub-threshold behavior of this denominator vs. triangle graph in Ref.~\cite{Jackura:2018xnx}, and the comparison of the Feynman denominator and time-ordered perturbation theory in Ref.~\cite{Zhang:2021hcl} where it was shown that the breaking of covariance in the latter is rather small. 

It should also be noted that there will be a much more complicated analytic structure in unphysical regions of the amplitude which $s$-channel unitarity alone cannot fix (analogous to two-body amplitudes). However, these structures are far away from the region in which we search for the $a_1$-pole and one can safely neglect them, i.e.~the expansion of the $C$-term in the Laurent series of Eq.~\eqref{eq:C-term} contains the dominant contributions.

The denominator of Eq.~\eqref{eq:Bterm} vanishes for any $x=\cos\theta\in[-1,1]$ according to the partial-wave decomposition of Eq.~\eqref{eq:BmatrixPartial}. For a fixed three-body energy $\sqrt{s}$ and incoming spectator momentum $p$ the singularities are given by
\begin{align}
p_\pm'=&\frac{px(p^2-\alpha^2)\pm\alpha\sqrt{\left(\beta+p^2\left(x^2-1\right)\right)^2-4m_\pi^2\beta}}{2\beta} \,,\nonumber \\
&\alpha(p)=\sqrt{s}-E_p \,,
~~
\beta(p,x)=\alpha^2(p)-p^2x^2 \,.
\label{eq:Bsingularities}
\end{align}
 The domain of real solutions is indicated in Fig.~\ref{fig:physmom} and bound by $p_\pm'(p,x=\pm 1)$.

\begin{figure}[t]
    \centering
    \includegraphics[width=\linewidth,trim=0 0 2.5cm 0,clip]{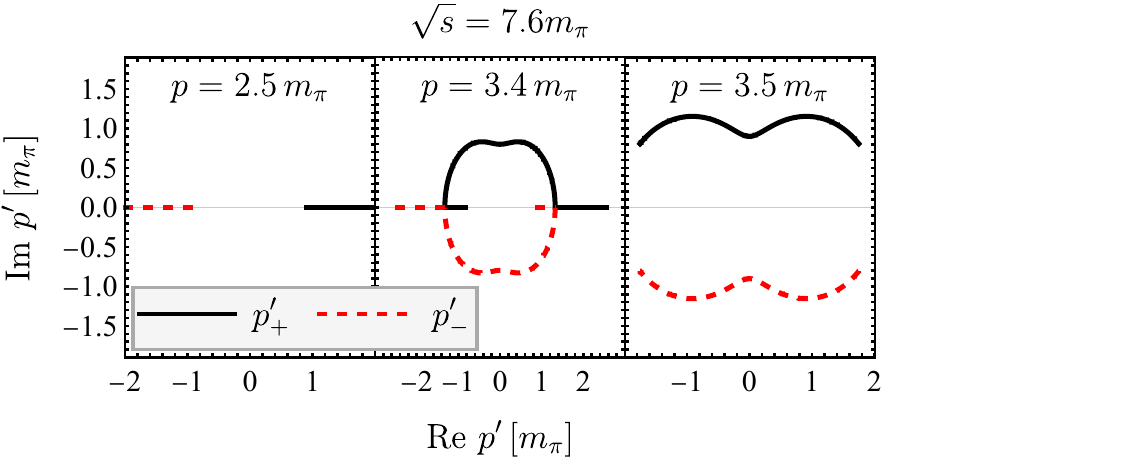}
    \caption{Position of three-body singularities in $p'(p,x)$ for fixed $p$, $\sqrt{s}=7.6\,m_\pi$, and $x\in [-1,1]$ for solutions $p_\pm'$ according to Eq.~\eqref{eq:Bsingularities}.
    }
    \label{fig:cplxmom}
\end{figure}

Not all solutions $p'$ are real, as Fig.~\ref{fig:cplxmom} shows. Notably, there are kinematic regions (e.g.~$p=3.4m_\pi$) in which the singularities fully enclose the origin which renders a naive integration from $0$ to $\infty$, or to any physically required cutoff $\Lambda$, impossible.  For the pions in the $\rho$ selfenergy in Eq.~\eqref{eq:tau-infinite} to be on-shell, the smallest physically required cutoff $p_\text{min}$ is given by the condition $\sigma(p)>4m_{\pi}^2$, this leads to

\begin{align}
p^2_\text{min}=\frac{9m_\pi^4-10m_\pi^2s+s^2}{4s} \,.
\label{eq:pmin}
\end{align}
Simultaneously, $p_\text{min}$ must be large enough to cover all physically allowed momenta in the pion-exchange, given by the extension of the domain shown in Fig.~\ref{fig:physmom}.  This can be determined through the vanishing argument of the square root of Eq.~\eqref{eq:Bsingularities} at $x=1$,
\begin{align}
\beta^2(p_\text{min},1)-4m_\pi^2\beta(p_\text{min},1)=0 \,.
\end{align}
The solution of this equation is also given by Eq.~\eqref{eq:pmin} as expected.

The crucial point is that the positions of the three-body singularities in $p'$ depend on the value of $p$ itself. For a suitably chosen contour SMC with $p\in\Gamma_{\rm SMC}$ and $p'\in\Gamma_{\rm SMC}$, the three-body cuts ``open up'' and allow for the integration of the scattering equation~\eqref{eq:Bethe-SalpeterPartial} which has been known for a long time~\cite{Haftel:1970zz}. See also Ref.~\cite{Doring:2013wka} for a similar numerical scheme in the context of Muskhelishvili-Omn\'{e}s equations.

\begin{figure}[t]
    \centering
    \includegraphics[width=\linewidth]{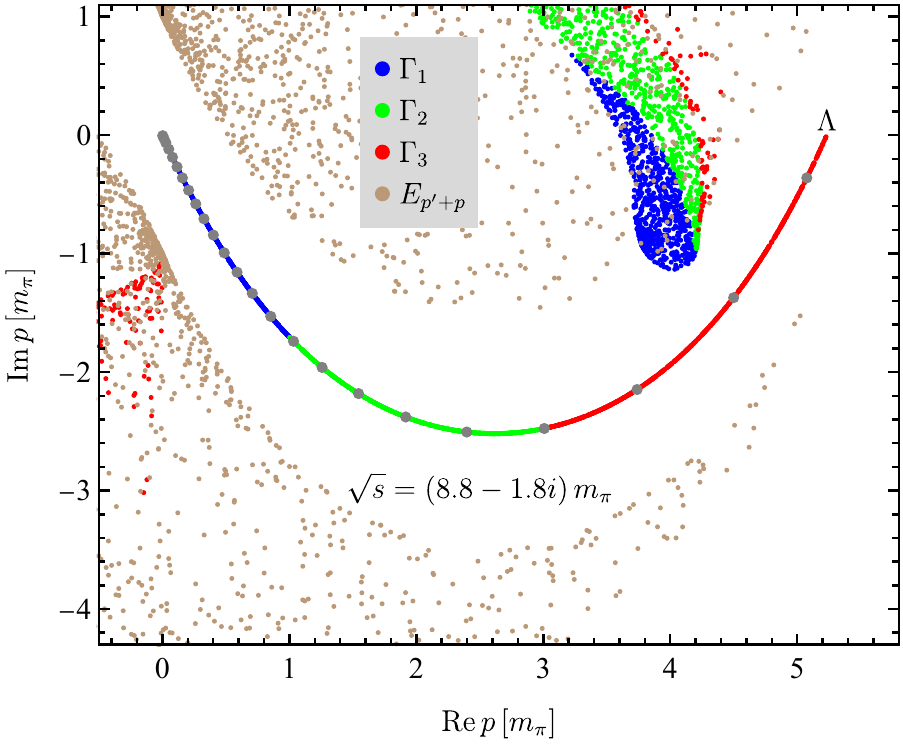}
    \caption{Three-body singularities at fixed, complex 3-body energy $\sqrt{s}$. The color-coded dots show both solutions  $p'$ of Eq.~\eqref{eq:Bsingularities} for different $p\in\Gamma_1$ (blue), $p\in\Gamma_2$ (green), $p\in\Gamma_3$ (red), and $x\in[-1,1]$. Solutions $p'(p,x)$ for $E_{p+p'}=0$ from Eq.~\eqref{eq:Bterm}, $p\in\Gamma_\text{SMC}$, and $x\in[-1,1]$ are indicated in light brown.
    }
    \label{fig:smc}
\end{figure}

The precise form of the SMC is not fixed. We choose the smooth contour depicted 
 in Fig.~\ref{fig:smc} that is split into different color-coded segments, $\Gamma_\text{SMC}=\bigcup_{i=1}^3\Gamma_i$, to show which parts of the SMC correspond to which positions of three-body singularities, indicated with the dot clouds for both $p_+$ and $p_-$. Additionally, the singularities $p'(p,x)$ of the $1/E_{p'+p}$ term in Eq.~\eqref{eq:Bterm} are indicated by brown dots irrespective of the value of $p\in\Gamma_\text{SMC}$. The SMC is parameterized as
\begin{align}
\Gamma_\text{SMC}=\{p|p=&t+iV_0(1-e^{-t/w})  \nonumber \\
&\times(1-e^{(t-\Lambda)/w}),\,t\in[0,\Lambda]\}\,.
\label{eq:SMC}
\end{align}
This expression contains a parameter for the initial and final slope, $w$, and another one for the extension of the SMC into the lower half-plane, $V_0$. In general, a larger $V_0$ allows one to go further into the complex $\sqrt{s}$ plane to look for poles; piecewise-straight contours are also possible, in general, but require more integration nodes than smooth paths for a given precision. An example of integration nodes is shown in Fig.~\ref{fig:smc} with the gray circles on top of $\Gamma_\text{SMC}$.

To avoid singularities of the $B$-term one simply ensures that $\Gamma_\text{SMC}$ never overlaps with the solutions of Eq.~\eqref{eq:Bsingularities},
\begin{align}
\Gamma_\text{SMC}\cap \{p'(p\in\Gamma_\text{SMC},x\in [-1,1])\}=\O \,;
\label{eq:cond1}
\end{align}
similarly, for the $E_{p'+p}$ term,
\begin{align}
\Gamma_\text{SMC}\cap\{p'|E_{p'+p}=0,p\in\Gamma_\text{SMC},x\in [-1,1]\}=\O\,.
\label{eq:cond2}
\end{align}
There is a region of $\sqrt{s}$ in the lower complex half-plane for which this is the case, and the extent of that region depends on $\Gamma_\text{SMC}$. We have made sure that with the SMC of Eq.~\eqref{eq:SMC} the corresponding $\sqrt{s}$-region covers the pole region of the $a_1(1260)$.

\subsection{Real and complex threshold openings}
\label{sec:complexbp}

In Sec.~\ref{sec:two-body} we have already discussed the analytic structure of the two-body amplitude, its (threshold) branch point at $\sigma=4m_\pi^2$, where the two Riemann sheets coincide, and the $\rho$ pole on the second Riemann sheet. We now discuss this amplitude in the presence of the SMC, i.e.~the two-body system being a subsystem of the three-body amplitude with the spectator momentum being on the SMC. 

Figure~\ref{fig:integrations} shows the SEC mapped to the $\sigma$ plane by $\sigma=4E_k^2$. It also shows the SMC mapped to this plane via Eq.~\eqref{eq:sigma}. This representation has the advantage that a crossing of SEC and SMC in the figure directly indicates a zero of the selfenergy denominator ($\sigma-4E_k^2$) of Eq.~\eqref{eq:tau-infinite} that has to be avoided,
\begin{align}
\sigma(p^2)\neq 4E_k^2~\forall~ p\in\Gamma_\text{SMC}\wedge k\in \Gamma_\text{SEC} \,.
\label{eq:nocrossing}
\end{align}
The last condition is that neither contour can cross the $\rho$ pole at $\sigma_\rho$, except for $p=0$,
\begin{align}
\sigma(p^2)\neq \sigma_\rho~\forall p\in\Gamma_\text{SMC}\backslash\{0\}~\wedge~4E_k^2\neq \sigma_\rho~\forall k\in \Gamma_\text{SEC}\,.
\label{eq:norho}
\end{align}
The conditions~(\ref{eq:cond0}, \ref{eq:cond1}-\ref{eq:norho}) constitute the complete set of rules to access all Riemann sheets in the problem. 

\begin{figure}[t]
    \centering
    \includegraphics[width=1.\linewidth]{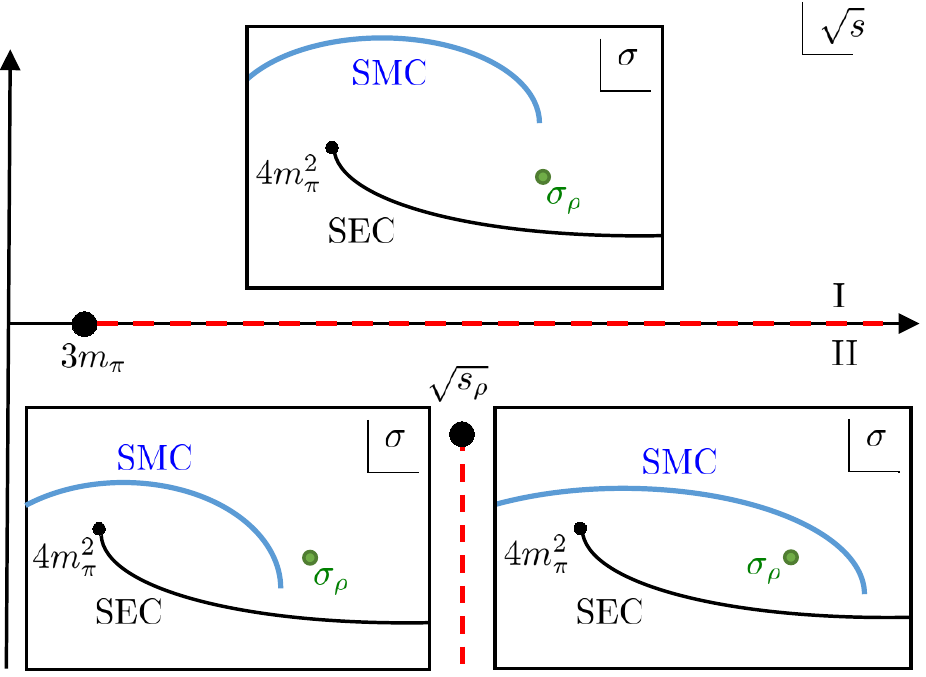}
    \caption{Analytic structure of the 3-body amplitude in the $\sqrt{s}$ plane. The real and complex branch points (thick black dots) are shown together with their respective cuts (red dashed lines). The $\pi\rho$ branch points lie on sheet II and induce additional sheets III and IV (not shown). The insets show the SMC mapped to the $\sigma$ plane (blue lines), in a qualitative way. Its position changes according to the approximate value of $\sqrt{s}$ where the insets are placed. In the $\sigma$ plane, the SEC (black lines) starting at the $\pi\pi$ threshold (small black dots) does not change if $\sqrt{s}$ changes, but the SMC does.
    }
    \label{fig:structure}
\end{figure}

The exclusion of $p=0$ in Eq.~\eqref{eq:norho} can be understood in the context of branch points. According to Eq.~\eqref{eq:sigma}, the condition $p=0$ and $\sigma=\sigma_\rho$ corresponds to $\sqrt{s}=\pm\sqrt{\sigma_\rho}+m_\pi$. In other words, at these complex three-body energies the spectator momentum integration starts at the $\rho$ pole. According to Sec.~\ref{sec:two-body}, if an integration limit coincides with a singularity, a branch point is generated. Therefore, taking only the square root of interest (positive $\text{Re}\,\sqrt{\sigma_\rho}$) and invoking the Schwarz reflection principle, we conclude that the three-body amplitude has branch points at $\sqrt{s}=\sqrt{s_\rho}:=\sqrt{\sigma_\rho}+m_\pi$ and $\sqrt{s}=\sqrt{s_\rho}^{\,*}=\sqrt{\sigma_\rho}^{\,*}+m_\pi$. We refer to them as $\pi\rho$ branch points in the following.

There is a third branch point: the two-body threshold induces the real-valued 3-body threshold at $s=(3m_\pi)^2$ because at that energy and $p=0$, we have $\sigma=4m_\pi^2$ according to Eq.~\eqref{eq:sigma}. The spectator momentum integration starts at the two-body branch point, which induces another branch point in the three-body amplitude. In Ref.~\cite{Ceci:2011ae} additional properties of these branch points were discussed.
 
The overall analytic structure of the three-body amplitude of Eq.~\eqref{eq:Bethe-SalpeterPartial} is visualized in Fig.~\ref{fig:structure}. It shows the real branch point at $\sqrt{s}=3m_\pi$ with its associated cut chosen along the real $\sqrt{s}$ axis defining sheets I and II. Also, the figure shows one of the complex branch points at $\sqrt{s}=\sqrt{s_\rho}$ which is situated on sheet II. The cut associated with the complex branch point is conveniently run into the negative imaginary $\sqrt{s}$-direction so that the shown Riemann sheet is the region closest to the physical axis. If regions behind that cut ought to be explored (defined as sheet III), more complicated contours must be chosen~\cite{Doring:2009yv}.

In addition, the insets in Fig.~\ref{fig:structure} show the $\sigma$ plane with the SMC and SEC similar as in Fig.~\ref{fig:integrations}. The position of the insets in the $\sqrt{s}$ plane qualitatively corresponds to the $\sqrt{s}$ used to map the SMC to the $\sigma$ plane, according to Eq.~\eqref{eq:sigma}. Note how the position of the SMC changes relative to the $\rho$ pole at $\sigma_\rho$. For example, for $\sqrt{s}$ to the left (right) of the $\pi\rho$ branch cut, the SMC passes the $\rho$ pole to the left (right). 
 
\begin{figure}[t]
    \centering
    \includegraphics[width=1.\linewidth]{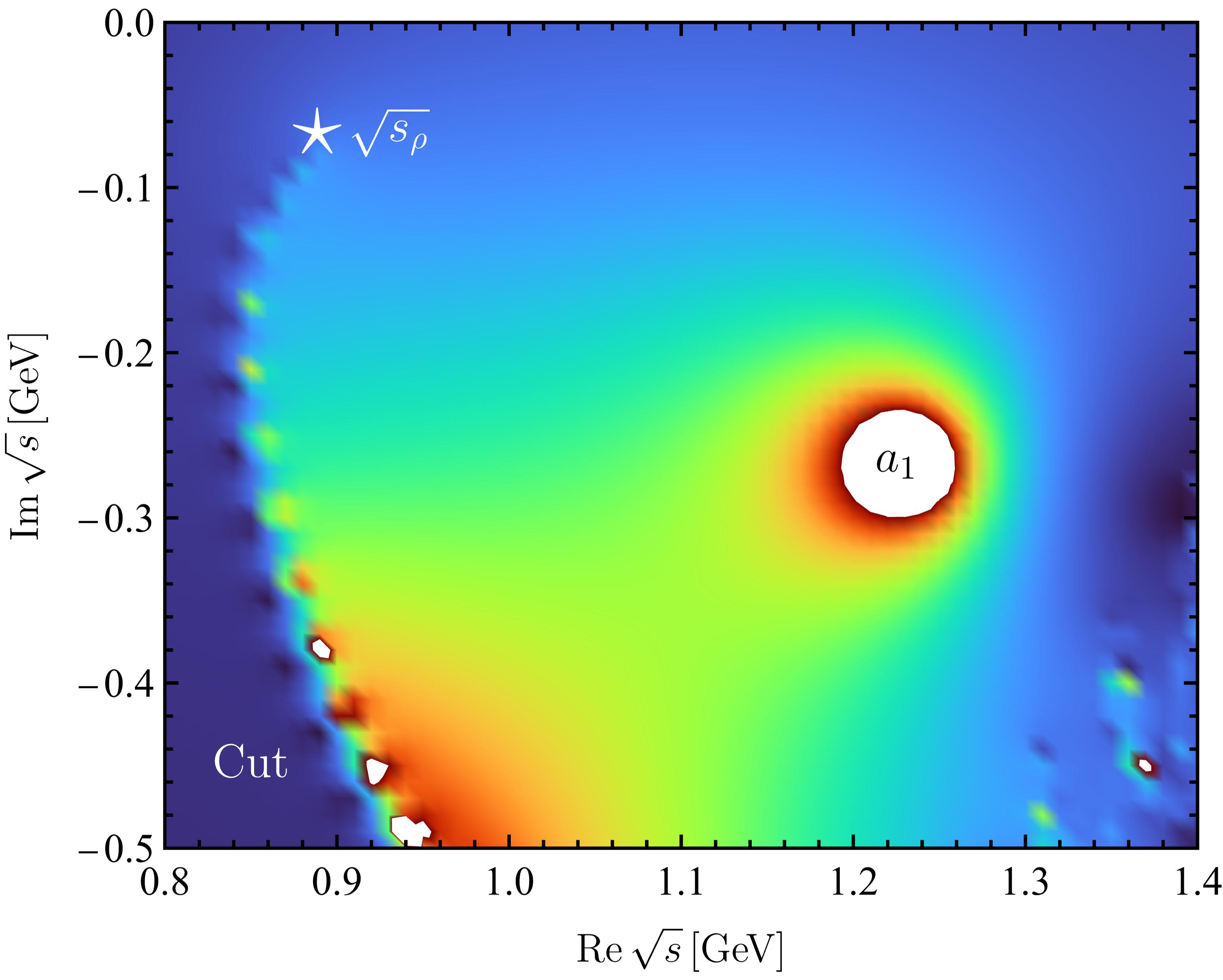}
    \caption{Typical amplitude $|T_{00}|$ of Eq.~\eqref{eq:Bethe-SalpeterPartial}, color coded from small values (dark blue) to large values (red; maxed-out values at white). The $a_1(1260)$ pole, $\pi\rho$ branch point at $\sqrt{s}=\sqrt{s_\rho}$ and its associated cut are also indicated. See text for further explanations.
    }
    \label{fig:complex_s}
\end{figure}
In Fig.~\ref{fig:complex_s} we show a typical picture of $|T_{00}|$ of Eq.~\eqref{eq:Bethe-SalpeterPartial} with the integration contours defined in Eqs.~\eqref{eq:SEC} and \eqref{eq:SMC}. The shape parameters that allow access to a sufficiently large region in the broad vicinity of the $a_1(1260)$ pole, which make the cut of the $\pi\rho$ branch point run approximately in the negative-imaginary $\sqrt{s}$ direction, are given in Table~\ref{tab:shape_parms}.
\begin{table}[tbh]
\begin{ruledtabular}
\begin{tabular}{llll}
\rule[-1.2ex]{0pt}{0pt}
$c_1\,[m_\pi]$  & $c_2\,[m_\pi^{-1}]$   & $w\,[m_\pi]$  & $V_0\,[m_\pi]$\\
\hline
\rule{0pt}{2.5ex} 
-7.16               & 0.418                     & 1.433           & -3.58            \\
\end{tabular}
\end{ruledtabular}
\caption{Shape parameters for the SEC in Eq.~\eqref{eq:SEC} and SMC in Eq.~\eqref{eq:SMC}.}
\label{tab:shape_parms}
\end{table}
The $a_1$ pole is always to the lower right of the $\pi\rho$ branch point in the $\sqrt{s}$ plane as Fig.~\ref{fig:complex_s} shows. Therefore, the qualitative positions of SEC and SMC in the $\sigma$ plane, corresponding to $\sqrt{s}$ taking the value of the $a_1$ pole, are given by the lower right inset of Fig.~\ref{fig:structure} which is also the situation shown in Fig.~\ref{fig:integrations}. Similar to Fig.~\ref{fig:integrations}, the cut induced by $\sqrt{s_\rho}$ is approximated by a series of poles due to the numerical discretizations, as Fig.~\ref{fig:complex_s} shows. While in the former case this was due to the self-energy integration, in the latter case it is due to the integration over the spectator momentum.

While Fig.~\ref{fig:complex_s} and all results in this paper have been obtained using the shape parameters of Table~\ref{tab:shape_parms}, the figure also shows that this choice is not universally valid for all three-body energies. In the lower right-hand corner (highest energies, farthest into the complex plane), we observe numerical fluctuations. These are poles induced by three-body singularities coinciding with the SMC as illustrated in Fig.~\ref{fig:smc}; they correspond to violations of Eqs.~\eqref{eq:cond1} or \eqref{eq:cond2}. If the analytic continuation in such regions of $\sqrt{s}$ is desired, one needs to choose a different SMC.

\begin{table*}[t]
\begin{ruledtabular}
    \begin{tabular}{c|llll|l|lll}
    $\Lambda$ [GeV] & $+0.73$ & $+0.90$ & $+1.05$ & $+1.2$  & $+0.73$ (no $B$)\B  & +0.73 (\cite{Protopopescu:1973sh} data) & +0.73 (\cite{Estabrooks:1974vu} data)
    & +0.73 (\cite{Pelaez:2019eqa} input)\\
     \hline
       Re~$\sqrt{s_0} $ [MeV] & $+1232^{+15}_{-<1}$ & $+1223$ & $+1231$ & $+1240$ & $+1174$ &$+1233$& $+1230 $ & $+1226$ \T\B \\
        Im~$\sqrt{s_0}$ [MeV]  & $-266^{+<1}_{-22}$ & $-269$ & $-244$ & $-251$ & $-252$&
        $-278$& $-261$ &$-253$  \B \\ \hline 
        $\chi^2/(65-6)$  & \phantom{+}$0.99$ & \phantom{+}$1.32$ & \phantom{+}$1.60$ & \phantom{+}$1.90$ &  \phantom{+}$2.56$ & \phantom{+}$0.99$ & \phantom{+}$0.98$ & 1.09 \T\B \\ \hline
        $c_{00}^{-1}$ & $+16.48^{+0.005}_{-0.007}$ & $+14.59$  & $+12.67$ & $+11.53$ &  $+20.16$ 
        & $+16.74$ & $+16.49$ & +15.83\T\B \\
         $c_{00}^0$ & $+1.729^{+0.008}_{-0.005}$ & $+1.750$ & $+1.843$ & $+2.073$ &  $+0.019$
         &$+1.712$& $+1.720$ & +2.077
         \B
         \\
         $m_{a_1}$ [GeV] & $+1.293^{+0.001}_{-0.000}$& $+1.287$ & $+1.281$ & $+1.278$ & $+1.391$
         &
         $+1.296$ & $+1.294$ & +1.324 \B
         \\
         $D_{f0}\times10^{7}$ [a.u.] & $-1.841^{+0.049}_{-0.027}$ & $-2.371$ &  $-2.126$ & $-2.250$ &$-0.925$
         &$-1.887$ & $-1.829$ & $-2.002$ \B\\
         $D_{f2}\times 10^{8}$  [a.u.]&  $+6.462^{+0.451}_{-0.149}$ & $+3.094$ & $+1.567$  & $+0.837$ & $-6.824$
         &$+6.718$ & $+6.512$ & $+2.073$ \B
         \\
         $D_{\tilde{f}}\times 10^{6}$  [a.u.]& $-1.319^{+0.002}_{-0.000}$ & $-1.358$  & $-1.338$ & $+1.372$ & $-1.235$
         &$-1.329$ & $-1.318$ & $-1.366$ \\
    \end{tabular}
    \caption{Pole positions $\sqrt{s_0}$ of the $a_1(1260)$, $\chi^2$, and fit parameters. Statistical uncertainties are only quoted for the first column.  The column labeled ``no $B$'' shows a fit for which we set the pion exchange term $B=0$. 
    The last three columns show variation from different two-body input as referenced in the labels. 
    The $c$-terms are unitless while 
    the abbreviation ``a.u.'' for the $D$-terms stands for ``arbitrary units'' because they contain the factor that connects to the un-normalized line shape data.
    }
    \label{tab:fitresults}
\end{ruledtabular}
\end{table*}

\begin{figure}[t]
\includegraphics[width=\linewidth,trim=0.1cm 1.6cm 0 0,clip]{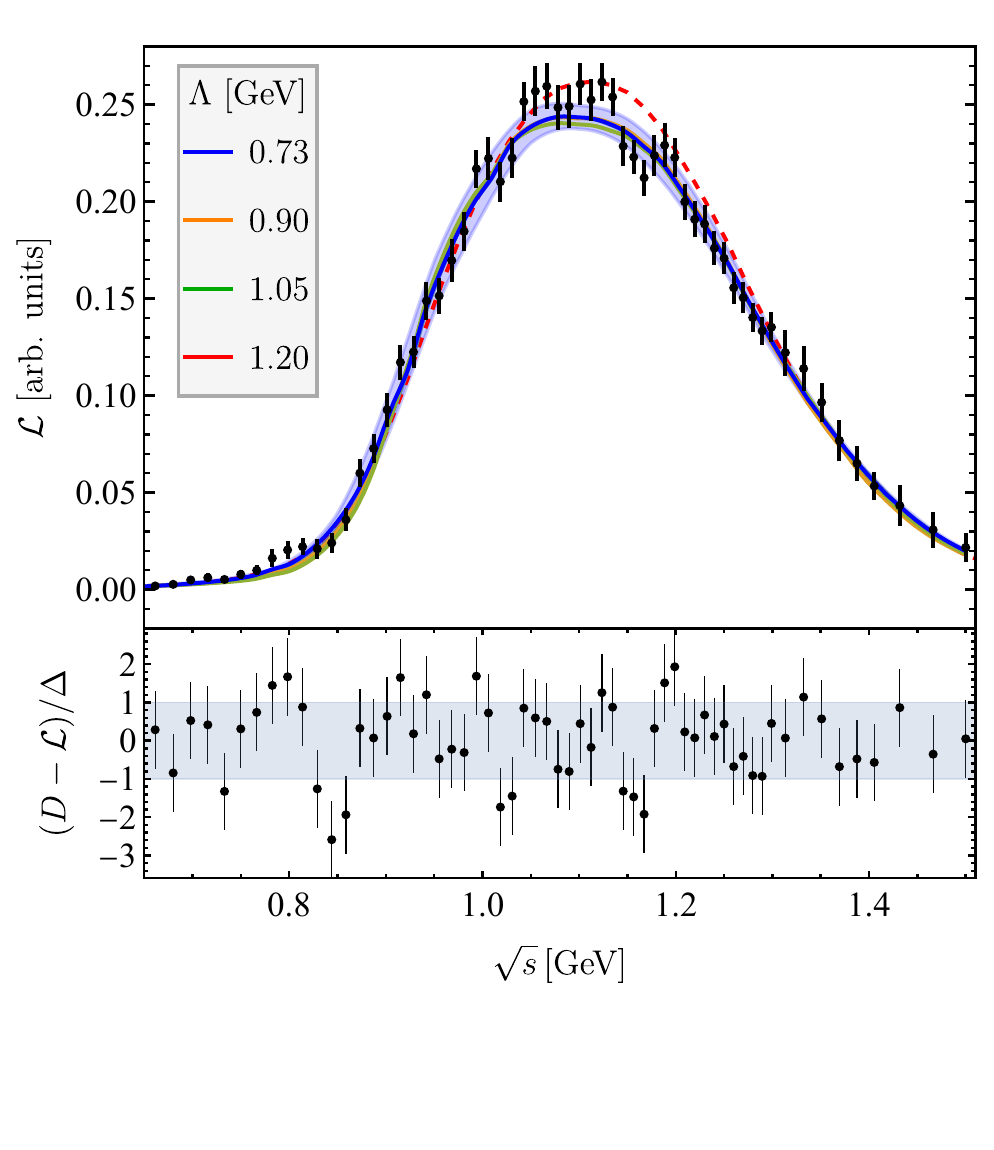}
\caption{
Fit to the line shape data from the ALEPH experiment~\cite{Davier:2013sfa} for different cutoff values (top) and the normalized residuals for the $\Lambda=0.73$~GeV case (bottom), where $D-\mathcal{L}$ is a given residual and $\Delta$ the pertinent data uncertainty. In the upper figure, the blue band shows the statistical uncertainties of the $\Lambda=0.73$ GeV fit, multiplied by ten for visibility. The red dashed line shows the $\Lambda=0.73$ GeV fit with all D-wave terms set to 0. 
}
\label{fig:phantom}
\end{figure}

\section{Results}
\label{sec:fit}

\subsection{Fit}

The free parameters of the model are fixed by a fit to the lineshape for the decay $\tau^- \rightarrow \pi^-\pi^-\pi^+\nu_\tau$. Data for this process measured in the ALEPH experiment were originally published in Ref.~\cite{Schael:2005am}. In Ref.~\cite{Davier:2013sfa} the unfolding method was improved and an error was fixed  (see  Ref.~\cite{websitealeph} for numerical values) . The data include  correlations that correspond to both systematic and statistical uncertainty. However, the systematic uncertainties are small relative to the statistical uncertainties, thus we neglect them. The $\chi^2$ can then be calculated with the formula
\begin{align}
\chi^2=(\vec{\mathcal{L}}-\vec{D})^T\Sigma^{-1}(\vec{\mathcal{L}}-\vec{D}) \ ,
\label{eq:fullchisquare}
\end{align}
where $\Sigma$ is the data covariance matrix, $\vec{D}$ is a vector containing the central values of the ALEPH line shape data and $\vec{\mathcal{L}}$ is a vector containing our prediction of the lineshape calculated with Eq.~\eqref{eq:lineshape} as a function of $\sqrt{s}$ at each of the central values of energy for the ALEPH data. We fit the 65 data points in the range $0.55$~GeV $<\sqrt{s}< 1.50$~GeV but do not include all correlations in $\Sigma$; instead we set all correlations to 0 except those between nearest neighbors. This choice is justified in Appendix~\ref{app:b}.

We fit the parameters $c_{00}^{-1}$ and $c_{00}^0$  from the expansion of the three-body term in Eq.~\eqref{eq:C-term}. We do not include $c_{00}^{1}$ or any higher $c_{00}$ terms and we do include any  $c_{10}$, $c_{01}$, or $c_{11}$ terms because including these terms in the fit when $\Lambda=0.73$ GeV increases the $\chi^2_\text{dof}$. Fits for other values of $\Lambda$ just serve to assess systematic effects and we do not try to change their parameterization.
Thus, our fit of the lineshape has a total of 6 free parameters: $c_{00}^{-1}$, $c_{00}^0$, and $m_{a_1}$ from Eq.~\eqref{eq:C-term}, and $D_{f0}$, $D_{f2}$, and $D_{\tilde{f}}$ from Eq.~\eqref{eq:firstdecayvertexprime}.

The line shape depends on $\Lambda$ in Eqs.~\eqref{eq:Gammabrev} and \eqref{eq:Bethe-SalpeterPartial}. A cutoff of $\Lambda=0.73$~GeV is the lowest possible value allowed by Eq.~\eqref{eq:pmin} if an upper limit of $\sqrt{s}=1.5$~GeV is chosen for the fit. We consider the case $\Lambda=0.73$~GeV to be our primary fit because it leads to the best $\chi^2$ as shown in Table~\ref{tab:fitresults}. However, to study systematic effects, we also vary $\Lambda$, leaving the two-body input encoded in the parameters $a_0$ and $a_1$ unchanged, and perform several fits. We list the pole position and free parameters of these fits in  Tab.~\ref{tab:fitresults}. As we increase $\Lambda$, $m_{a_1}$ and $c_{00}^{-1}$ decrease, whereas $c_{00}^0$  $m_{a_1}$ increases. The pole position remains relatively unchanged, indicating the physical pole position does not depend on the cutoff.

\begin{figure*}[t]
\centering
\includegraphics[height=8.2cm]{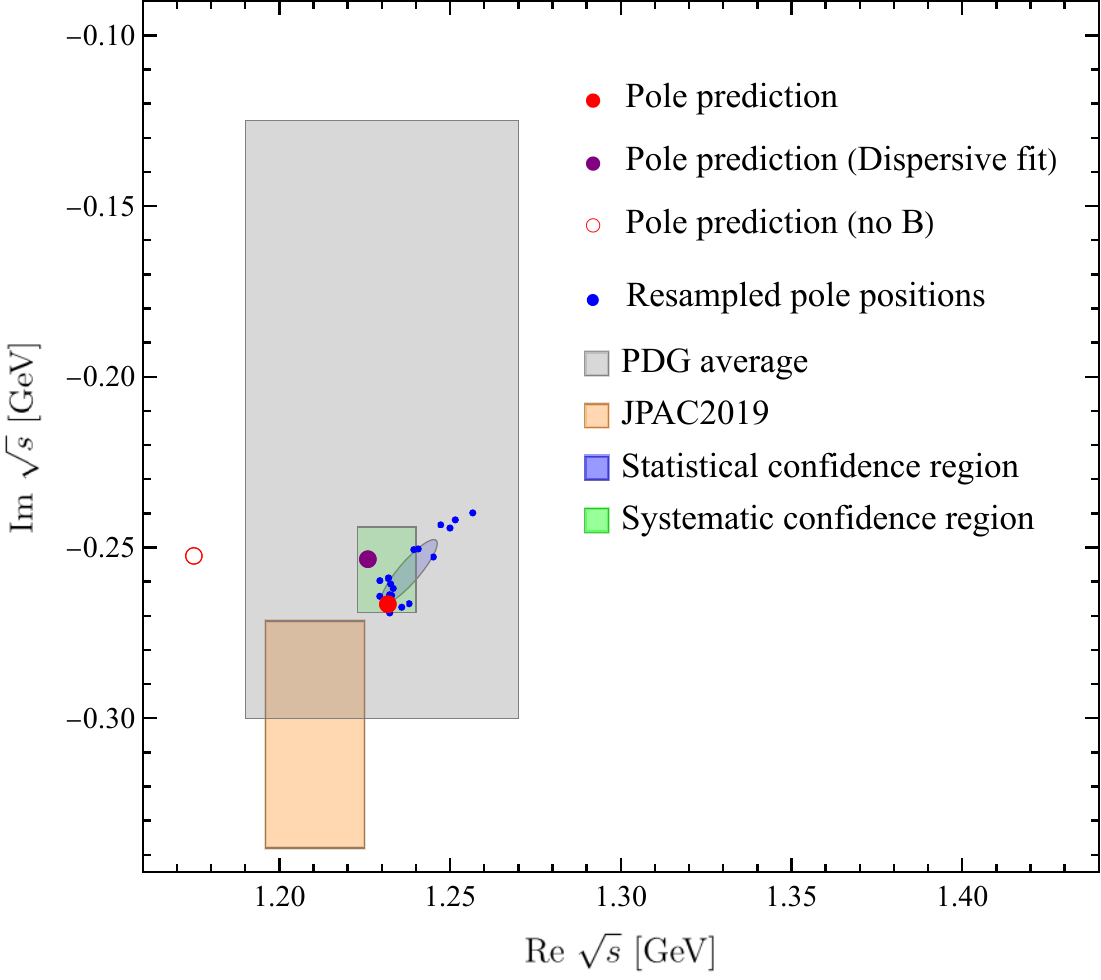}
~~~~~
\includegraphics[height=8.2cm]{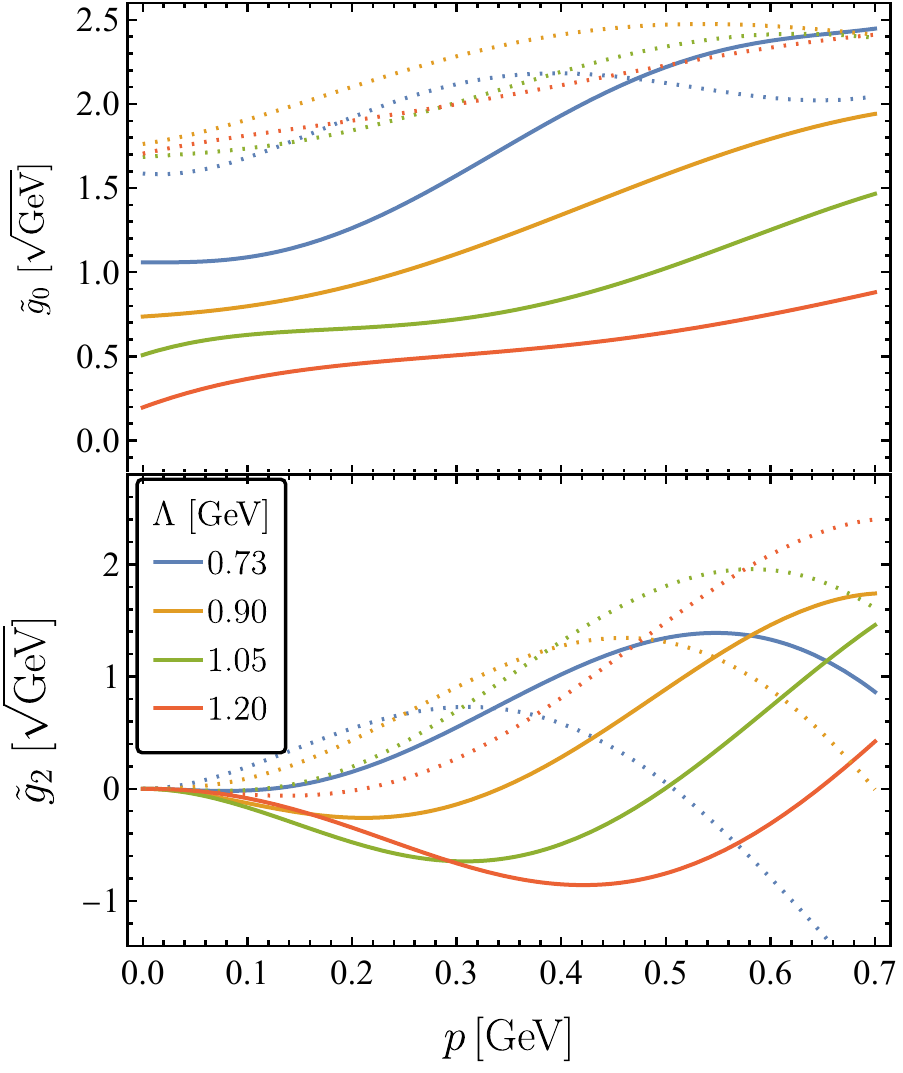}
\caption{
Left: Compilation of pole positions determined in this work including statistical and systematic uncertainties. For convenience, the PDG~\cite{ParticleDataGroup:2020ssz} average and a result by JPAC~\cite{JPAC:2018zwp} are quoted as well. See text for further explanations. 
Right: Pole couplings according to Eq.~\eqref{eq:couplings} as a function of a real spectator momentum $p$. These quantities play the role of spectator-momentum dependent branching ratios. Results for S-wave are shown at the top and for D-wave at the bottom, for different cutoffs $\Lambda$ as indicated. The solid (dashed) lines show the real (imaginary) parts.}
\label{fig:A1-pole}
\end{figure*}

\subsection{Discussion}
\label{sec:discussion}

The $a_1(1260)$ lineshape data and best fits for different cutoffs are shown in Fig.~\ref{fig:phantom} with the solid lines. Fit uncertainties are indicated with the blue band for our main result ($\Lambda=0.73$~GeV). For better visibility, the band width is multiplied by a factor of 10. The plot of reduced residuals for the $\Lambda=0.73$~GeV case (bottom of Fig.~\ref{fig:phantom}) shows that there are no obvious systematic deviations of the fit from data, except maybe for a structure at $\sqrt{s}\approx 0.8$~GeV. Also,  around that energy the fits for different $\Lambda$ differ from each other more than in other energy regions, and a substantial fraction of the $\chi^2$ for larger cutoffs $\Lambda$ (see Table~\ref{tab:fitresults}) arises in this energy region. 
In general, the pertinent fits are all very close together which is  reflected in the very small variation in pole positions indicated in Table~\ref{tab:fitresults}.

We show with a red dashed line in Fig.~\ref{fig:phantom} the line shape with $\Lambda=0.73$~GeV calculated with the D-wave contributions set to 0. The difference in this case from the blue line shows the contributions of the D-wave term. This difference is small, which is qualitatively in line with the PDG~\cite{ParticleDataGroup:2020ssz}. Note that this is a prediction at this point; the D-wave can only be determined more quantitatively once Dalitz plots are analyzed, as measured, e.g.~by CLEO~\cite{CLEO:1999rzk}.

We test the impact of the rescattering term by performing a fit in which we exclude 
$B^J_{LL'}(q_1,p)$ from Eq.~\eqref{eq:Bethe-SalpeterPartial}. This pion-exchange term is required by three-body unitarity~\cite{Mai:2017vot} and omitted in Ref.~\cite{JPAC:2018zwp}. As Table~\ref{tab:fitresults} (last row) and Fig.~\ref{fig:A1-pole} (open red circle) show, the $a_1(1260)$ pole position is significantly shifted in the refit without the B-term. Still, Fig.~\ref{fig:A1-pole} also shows that, even without the $B$-term, our model is not identical to the one of JPAC~\cite{JPAC:2018zwp} which might be due to a slightly different treatment of the two-body input, different cutoffs, or the fact that in Ref.~\cite{JPAC:2018zwp} the lineshape data of Ref.~\cite{Schael:2005am} is fitted, while we fit the data of Ref.~\cite{Davier:2013sfa}.

We show our pole predictions in Fig.~\ref{fig:A1-pole}. 
Statistical uncertainties are calculated through a re-sampling procedure (blue dots) in a two-step process.  Firstly, the two-body data from Refs.~\cite{Protopopescu:1973sh,Estabrooks:1974vu} are resampled 20 times with a normal distribution using the given data uncertainties. Parameters $a_0$ and $a_1$ (given in Eq.~\eqref{eq:Kmatrix}) are fit to each resampled set. Secondly, 20 sets of resampled ALEPH data~\cite{Davier:2013sfa, websitealeph} are generated. A fit is performed for each resampled set using the different values of $a_0$ and $a_1$ calculated in the first step. From each of these fits, a pole position is calculated, shown with blue dots in Fig.~\ref{fig:A1-pole}.  We also show the pertinent error ellipse keeping in mind that this is a non-linear fit problem.

The green rectangle in Fig.~\ref{fig:A1-pole} shows the region of pole positions from different cutoffs $\Lambda$ according to Table~\ref{tab:fitresults}. To that we add a second source of systematic uncertainties from the different two-body input according to the last two columns of Table~\ref{tab:fitresults}. Such variations in $\Lambda$ help assess the influence of inherent model uncertainties so we add then as systematic error to our result for the $a_1(1260)$ pole position quoted in the next section.
Our entire confidence region, including both statistical and systematic uncertainties, lies entirely within the PDG estimate of the $a_1(1260)$ denoted with the gray rectangle, and it is not in strong tension with the JPAC result~\cite{JPAC:2018zwp} (orange rectangle). 

The purple dot in Fig.~\ref{fig:A1-pole} shows the predicted pole position using the improved two-body input from Ref.~\cite{Pelaez:2019eqa}, see Sec.~\ref{sec:twobody}. The value of $\sqrt{s_0}=(1226-253i)$~MeV lies within our systematic uncertainty which demonstrates that there is not a very strong dependence on the two-body input.

We extract the residues of the pole position using the coupled-channel partial-wave amplitude of Eq.~\eqref{eq:Bethe-SalpeterPartial}. It can be expanded in $\sqrt{s}$ around the pole position $\sqrt{s_0}$ of the $a_1(1260)$,
\begin{align}
T^J_{LL'} (p,p')=\frac{\tilde g_{L}(p)\, \tilde g_{L'}(p')}{\sqrt{s}-\sqrt{s_0}}+{\cal O}(1) \,,
\label{eq:couplings}
\end{align}
with $\tilde g_L$ playing the role of (Breit-Wigner) branching ratios, but defined at the pole~\cite{ParticleDataGroup:2020ssz}. In addition, for the current case of $\pi\rho$ scattering in S and D-waves, the $\tilde g$ are necessarily functions of spectator momentum, $\tilde g_L\equiv\tilde g_L(p)$. Analogously one might think of resonance transition form factors that are closely related to pole residues depending on photon virtuality~\cite{Kamano:2018sfb, Mai:2021vsw}. For a numerically stable method to calculate residues see Appendix C of Ref.~\cite{Doring:2010ap}. We show the $\tilde g_L(p)$ for real spectator momenta $p$ in Fig.~\ref{fig:A1-pole}, which requires another analytic extrapolation from the complex $p$ on the spectator momentum contour (SMC) at which the solution is calculated.


As Fig.~\ref{fig:A1-pole} shows, the $a_1$ resonance does couple to the $\pi\rho$ D-wave channel even if the corresponding coupling term appearing in Eq.~\eqref{eq:firstdecayvertexprime} is not fitted, $c_{22}^{-1}=0$, and, similarly, $c_{20}^{-1}=c_{02}^{-1}=0$ in Eq.~\eqref{eq:C-term}. This is due to the $B$-term which always allows for non-diagonal transitions between S and D-wave channels. The D-wave decay is clearly smaller than the S-wave decay, and the contribution to the lineshape from D-wave (at real energies) is very small, see Fig.~\ref{fig:phantom}.
While our complex pole couplings are a prediction at this point, in future work they can be tested and even extracted from data by analyzing Dalitz plots of the $a_1$ decay such as measured at CLEO~\cite{CLEO:1999rzk}.

\section{Conclusions}

In this work we  have detailed how the pole position of the $a_1(1260)$ meson can be determined using a manifestly unitary three-body formalism. The three-body dynamics of the decay are fully taken into account, including the line shape corrections due to pion exchange (sometimes referred to as ``rearrangement`` graph). This process is a direct consequence of unitarity. It ensures that, apart from the usual isobar-spectator propagation in the $s$-channel, this is the only possible on-shell arrangement of three pions. Also, the amplitude necessarily exhibits two independent integrations that cannot be simply recast and factorized into the phase space calculation.

Three-body cuts and the two integrations imply problems for the analytic continuation of the amplitude to the complex pole position of the $a_1(1260)$. We explain in detail how the continuation is achieved by contour deformation and how different Riemann sheets are induced by an appropriate choice of integration contours.

Upon implementation, we find that the pion exchange term does have significant influence on the pole position of the $a_1(1260)$; taking into account nearest-neighbor correlations in the data from ALEPH~\cite{Davier:2013sfa}, the pole position is determined to be 
\begin{align}
\sqrt{s_0}=(1232^{+15+9}_{-0-11}-i266^{+0+15}_{-22-27})~\text{MeV} \, ,
\end{align}
where the first errors are statistical (including nearest-neighbor correlations) and the second are systematic.  The systematic uncertainties stem from the cut-off dependence and two-body input as shown in Table~\ref{tab:fitresults}. 

The current calculation is restricted to $\rho\pi$ channels in S and D-wave.  Future upgrades to include more coupled channels, like sub-dominant $\sigma\pi$, will enable accurate simultaneous fits to line shape and Dalitz plot data to exploit unitarity which relates them.



\bigskip

\textit{Acknowledgments} ---
This material is based upon work supported by the National Science Foundation under Grant No. PHY-2012289 and the U.S. Department of Energy, Office of Science, Office of Nuclear Physics under Award Number DE-SC0016582, DE-AC05-06OR23177, and DE-FG02-95ER40907. Work of MM is supported in part by  the Deutsche Forschungsgemeinschaft (DFG, German Research Foundation) and the NSFC through the funds provided to  the Sino-German Collaborative  Research  Center  TRR110  ``Symmetries and  the Emergence of  Structure in  QCD'' (DFG  Project  ID 196253076  - TRR  110,  NSFC Grant  No.  12070131001).CC  is  supported by  UK  Research  and  Innovation  grant  MR/S015418/1.  HA thanks the Avicenna-Studienwerk e.V.\ for  financial support with funds from the BMBF.

\bigskip
\bibliography{BIB}

\clearpage
\appendix

\section{Technical details on spin-1 systems}
\label{app:a}

For each helicity state $\lambda\in\{-1,0,+1\}$ of the spin-1 field, the four-vector $\varepsilon$ depends on the direction of the propagation~\cite{Chung:1971ri} as
\begin{align}
\varepsilon_0(\bm{p})&=\frac{1}{m_\rho}\begin{pmatrix}p\\E^\rho_{p}
\cos\phi_{\bm{p}}\sin\theta_{\bm{p}}\\
E^\rho_{p}\sin\phi_{\bm{p}}\sin\theta_{\bm{p}}\\
E^\rho_{p}\cos\theta_{\bm{p}}
\end{pmatrix}\,,
\\
\varepsilon_{\pm 1}(\bm{p})&=\frac{1}{\sqrt{2}}\begin{pmatrix}
0 \\
\mp \cos{\theta_{{\bm{p}}}}\cos{\phi_{{\bm{p}}}}+i\sin{\phi_{{\bm{p}}}} \\
\mp \cos{\theta_{{\bm{p}}}}\sin{\phi_{{\bm{p}}}}-i\cos{\phi_{{\bm{p}}}} \\
\pm \sin{\theta_{{\bm{p}}}} \end{pmatrix}\,,
\label{eq:helicity-vectors}
\end{align}
where $E^\rho_p:=\sqrt{m_\rho^2+\bm{p}^2}$ and we chose $m_\rho=6.44\,m_\pi$. On-shell, this fulfills required properties, such as the transversality, i.e., $p_\mu\varepsilon^\mu=0$ exactly, see Ref. \cite{Chung:1971ri}. Away from the on-shell point one can generalize the above definitions using $m_\rho\to\sqrt{E_p^2-p^2}$. However, as the difference between both versions does not lead to new singularities of the spin-1 propagator, perturbation theory is viable, allowing one to reabsorb it into the local terms~\cite{Bruns:2013tja}.

Equation~\eqref{eq:rhos} requires the calculation of the helicity sum for the $s$-channel $\rho$ propagation,
\begin{align}
\sum_\lambda\epsilon_{\lambda,\mu}(\bm{p})\epsilon^*_{\lambda,\nu}(\bm{p})
=-g_{\mu\nu}+\frac{p_\mu p_\nu}{m_\rho^2} \,.
\label{eq:helicity-sum}
\end{align}

\begin{figure}[tb]
    \centering
    \includegraphics[width=\linewidth,trim=0 1.6cm 0 0,clip]{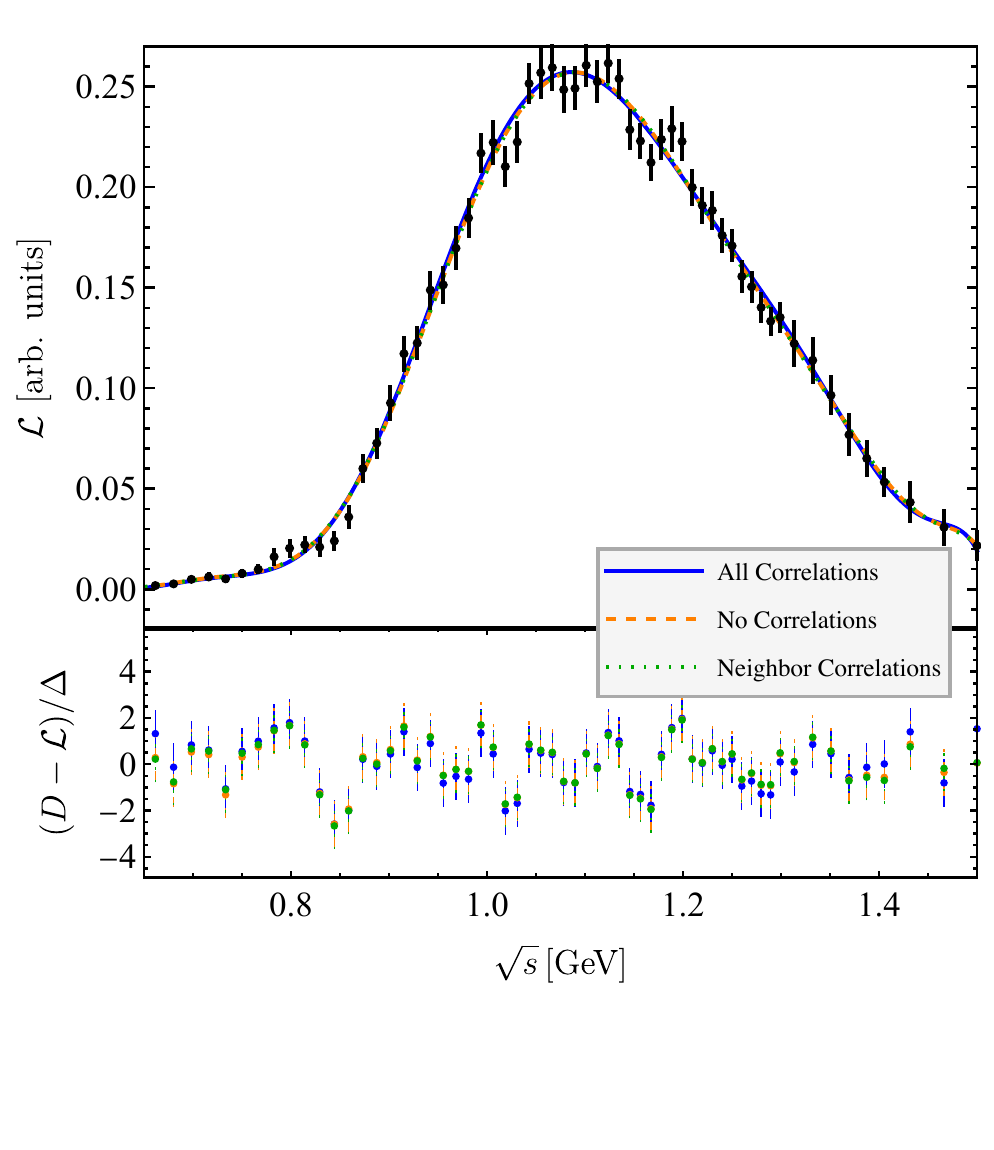}
    \caption{The fits (upper row) and normalized residuals (lower row) using a Legendre Polynomial Expansion to the ALEPH data ($n=10$ in Eq.~\eqref{eq:pheno}). The color coding indicates the cases where the covariance matrix includes all correlations (blue), no correlations (orange), and nearest-neighbor correlations (green).
    }
        \label{fig:phenomtest}
\end{figure}

\begin{table}[thb]
    \begin{ruledtabular}
    \begin{tabular}{lccc|c}
    & \rot{\begin{minipage}{1.8cm}\flushleft
    nearest\newline\noindent neighbor\\ correlations\end{minipage}} 
    & \rot{\begin{minipage}{1.5cm}\flushleft
    no \\ correlations\end{minipage}} 
    & \rot{\begin{minipage}{1.5cm}\flushleft
    all\\ correlations\end{minipage}} 
    & \rot{\begin{minipage}{1.8cm}\flushleft
    all corr.\\ (2005 data)\end{minipage}} \\
    \hline
    $\chi^2$  \T & 64  & 62 & 127 & 16 \hspace*{0.4cm}\\
    No. data \T & 63 & 63 & 63 & 75\hspace*{0.4cm}\\
    \end{tabular}
    \caption{Total $\chi^2$ for fits to the ALEPH data while various correlations were included.
    The first three entries refer to the 2013 data of Ref.~\cite{Davier:2013sfa} while the last column refers to the 2005 data of Ref.~\cite{Schael:2005am}.}
    \label{tab:my_label}
    \end{ruledtabular}
\end{table}

\begin{figure}[thb]
    \centering
    \includegraphics[width=\linewidth,trim=0 1.5cm 0 0 ,clip]{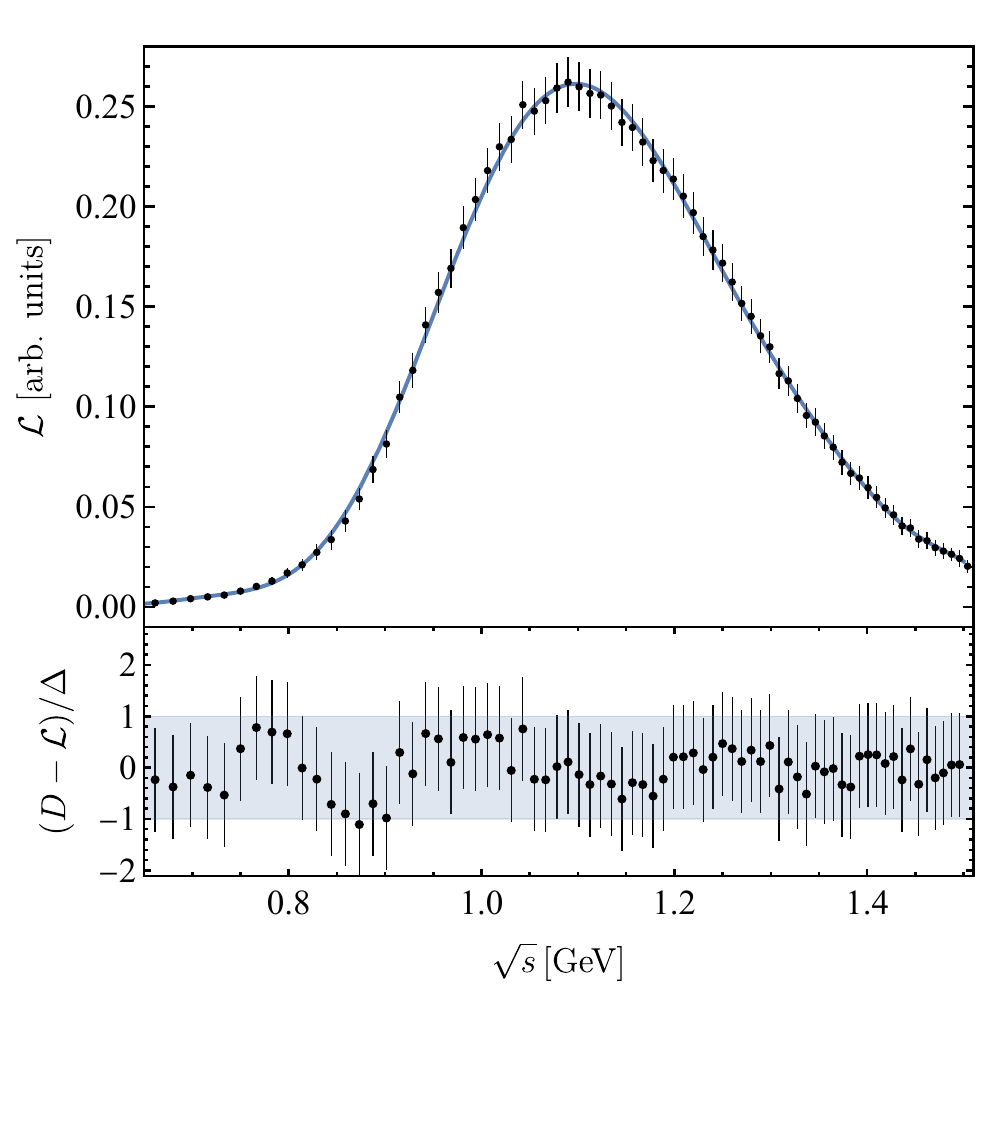}
    \caption{Older ALEPH data~\cite{Schael:2005am}, pertinent fit, and normalized fit residuals. These data are not used in the analysis. See Fig.~\ref{fig:phantom} for notation.
    }
    \label{fig:2005data}
\end{figure}

\section{Data consistency tests}
\label{app:b}

Rather than considering all data correlations from Ref.~\cite{Davier:2013sfa} in the covariance matrix for the calculation of $\chi^2$ using Eq.~\eqref{eq:fullchisquare}, we include only nearest-neighbor correlations. All other correlations are neglected because we find that no reasonably smooth curve can describe the data when they are included as shown in the following.

We fit the ALEPH data with a Legendre Polynomial expansion, $f(\sqrt{s},x_i)$ given by
\begin{align}
f(\sqrt{s},x_i)=\sum_{i=0}^n x_i P_n\left(\frac{\sqrt{s}-\sqrt{s_0}}{\sqrt{s_0}}\right),
\label{eq:pheno}
\end{align}
where $\sqrt{s_0}=1$~GeV is chosen such that the argument is always in the interval $[-1,1]$ and the $x_i$ are fitted by minimizing the $\chi^2$, 
\begin{align}
\chi^2=(\vec{f}(x_i)-\vec{\mathcal{L}})^T\Sigma^{-1}(\vec{f}(x_i)-\vec{\mathcal{L}}) \ ,
\label{eq:fullchisquareappendix}
\end{align}
where $\vec f$ is constructed from the data at $\sqrt{s_i}$, $(\vec f)_j=f(\sqrt{s_j},x_i)$. Similarly, the central values of the data are collected in the vector $\vec{\mathcal{L}}$.
When we fit in the range  $0.6~\text{GeV}<
\sqrt{s}<1.5$~GeV, with $n=10$ including all correlations we find that the minimum $\chi^2$ for the 63 data points is 127 ($\chi^2_\text{dof}=2.43$) as Table~\ref{tab:my_label} shows. Increasing the number of polynomials in this expansion will decrease the total $\chi^2$ but does not decrease the $\chi^2_\text{dof}$. For example, when $n=15$, $\chi^2= 122$ and, therefore,  $\chi^2_\text{dof}=2.58$. We show the $n=10$ fit and the residuals in  Fig.~\ref{fig:phenomtest}. These results indicate that it is very difficult for any smooth curve, regardless of whether or not it is theoretically justifed, to describe the data.

In contrast, when we use the same function to perform a fit where we include only the uncorrelated uncertainties and set all correlations to 0, we find the total $\chi^2$ for the 63 data points to be 62 indicating that the uncorrelated data can be easily fitted with a smooth function. The orange line and data of Fig.~\ref{fig:phenomtest} show the fit and residuals when none of the correlations are included in the data. This fit is quite similar to the one that includes all correlations even though the $\chi^2$ is very different. 

We note that while this case demonstrates that the uncorrelated data can be reasonably described by a smooth function, the residuals still display some noticeable correlation. In order to account for these correlations we introduce another case, shown in blue in Fig.~\ref{fig:phenomtest}. Here, we include only the nearest-neighbor correlations. When we fit the data with $n=10$ to this case, we obtain $\chi^2=64$. As this case includes the maximum of correlations that can be reconciled with a statistically sound description of the data, we regard this case as the data set for the analysis described in the main text.

We also apply this phenomenological test to the data as they were originally published in Ref.~\cite{Schael:2005am}. These data, shown in Fig.~\ref{fig:2005data}, have since been updated in Ref.~\cite{Davier:2013sfa}. We find that the older data are considerably over-fit for $n=10$  with a total $\chi^2$ of 16 as shown in Table~\ref{tab:my_label}. We therefore discard these data.

In summary, all three fits shown in Fig.~\ref{fig:phenomtest} are quite similar, even for the residuals. This implies that the best-fit parameters for each case will be quite similar regardless of which correlations are included. Thus, our choice of data (nearest-neighbor correlations only) affects the value of our $\chi^2$, but it does not have much effect on the best values for pole position or residues.

\end{document}